\documentclass[preprint2]{aastex6}

\shorttitle{GMOS Transmission Spectrum of WASP-4\normalfont{b}}
\shortauthors{Huitson et al.}

\begin{document}


\title{Gemini/GMOS Transmission Spectral Survey: Complete Optical Transmission Spectrum of the hot Jupiter WASP-4\MakeLowercase{b}}



\author{C.M.~Huitson\altaffilmark{1}, J-.M~D\'{e}sert\altaffilmark{2}, J.L.~Bean\altaffilmark{3}, J.J.~Fortney\altaffilmark{4}, K.B.~Stevenson\altaffilmark{5}, M. Bergmann\altaffilmark{6}}
\affil{$^1$CASA, University of Colorado, 389 UCB, Boulder, CO, 80309-0389, USA \\
$^2$API, University of Amsterdam, P.O. Box 94249, 1090 GE Amsterdam, The Netherlands \\
$^3$ Department of Astronomy and Astrophysics, University of Chicago, Chicago, IL 60637, USA \\
$^4$ Department of Astronomy and Astrophysics, University of California, Santa Cruz, CA 95064, USA \\
$^5$ Space Telescope Science Institute, 3700 San Martin Dr, Baltimore, MD 21218, USA \\
$^6$ NOAO \& Gemini Observatory, present address Palo Alto, CA}


\email{catherine.huitson@colorado.edu}

%
%
%
%



\begin{abstract}


We present the complete optical transmission spectrum of the hot Jupiter WASP-4b from 440-940~nm at $R \sim $ 400-1500 obtained with the Gemini Multi-Object Spectrometers (GMOS); this is the first result from a comparative exoplanetology survey program of close-in gas giants conducted with GMOS.
WASP-4b has an equilibrium temperature of 1700~K and is favorable to study in transmission due to a large scale height (370~km). 
We derive the transmission spectrum of WASP-4b using 4 transits observed with the MOS technique. We demonstrate repeatable results across multiple epochs with GMOS, and derive a combined transmission spectrum at a precision about twice above photon noise, which is roughly equal to to one atmospheric scale height.
The transmission spectrum is well fitted with a uniform opacity as a function of wavelength. The uniform opacity and absence of a Rayleigh slope from molecular hydrogen suggest that the atmosphere is dominated by clouds with condensate grain size of $\sim 1$~$\mu$m.
This result is consistent with previous observations of hot Jupiters since clouds have been seen in planets with similar equilibrium temperatures to WASP-4b.
We describe a custom pipeline that we have written to reduce GMOS time-series data of exoplanet transits, and present a thorough analysis of the dominant noise sources in GMOS, which primarily consist of wavelength- and time- dependent displacements of the spectra on the detector, mainly due to a lack of atmospheric dispersion correction. 

\end{abstract}




\section{Introduction}

Transiting planets allow us to study exoplanet atmospheres through transmission spectroscopy. A measurement of the wavelength-dependent flux dimming while the planet passes in front of its host star reveals characteristic signatures of atmospheric constituents \citep{seagersasselov00,charbonneau02}. Several hot Jupiters have now been studied using transmission spectroscopy, and their upper atmospheres show surprising diversity, with the causes of the variations not yet understood. 

The most prominent emerging difference across the known population is that some planets have optical spectra showing strong, wide alkali features predicted from studies of brown dwarfs \citep{seagersasselov00,charbonneau02,sing08,sing08b,2009ApJ...699..478D,2016ApJ...827...19F}, while other hot Jupiters have featureless optical transmission spectra or significant muting of predicted features, likely indicative of high altitude aerosol opacities \citep{pont08,pont13,2011A&A...526A..12D,sing11,sing12,sing13,huitson12,2014AJ....147..161S,2014MNRAS.437...46N,2015MNRAS.447..463N,2014Natur.505...69K}. Near-infrared transmission spectra show similar variations in molecular absorption (e.g. \citealt{deming13,huitson13,2013ApJ...779..128M,2014ApJ...791...55M,2014ApJ...793L..27K,2015ApJ...814...66K}).

The hot Jupiters that have been studied so far in transmission spectroscopy span a large range of equilibrium temperatures, from 1300 to 3000~K but there is no clear correlation between the presence of aerosols and equilibrium temperature \citep{2016Natur.529...59S}. The sample also covers a large range of host star type, stellar activity and planet surface gravity as well as there being differences in day-night thermal recirculation efficiencies. This picture is further complicated by the many factors that could affect aerosol formation such as photochemistry and night-side cold-trapping \citep{showman09,spiegel09,parmentier12,moses11,2016ApJ...817L..19T}.

Aside from the presence or absence of high-altitude aerosols, there are also differences in abundances between the various planets, with some showing prominent Na~\textsc{i} features and some showing prominent K~\textsc{i} features with negligible Na~\textsc{i} signal (e.g. \citealt{2014MNRAS.437...46N,2015MNRAS.446.2428S}). Since K~\textsc{i} condenses at a temperature hotter than Na~\textsc{i} and also ionizes more readily it is difficult to see a mechanism for such differences if the planets have a similar formation history.

The best way to proceed to better understand the properties of hot Jupiter atmospheres is to conduct large-scale studies of several planets through survey programs. Such studies enable comparative observation of multiple targets as well as ensuring a consistent methodology across the sample. For this reason, we initiated a 3.5-year survey of 9 hot Jupiters using the ground based Gemini telescopes (P.I.~ J-.M. D\'{e}sert, program numbers in Section~\ref{sec_obs}). 

We conducted observations in the optical using the MOS technique applied to transiting exoplanets \citep{2010Natur.468..669B} with the Gemini Multi-Object Spectrometers (GMOS) \citep{2004PASP..116..425H,2002PASP..114..892A} installed on both Gemini North and South telescopes to achieve a near full sky coverage with a similar instrumental setup.
The goal of this survey is to search for the dominant atmospheric absorbers, constrain abundances of alkali metals and detect the presence or absence of aerosols. We selected our planets so that they sample a range of masses, radii and host star types.

We present in this paper the first result from this survey: the complete optical transmission spectrum of the hot Jupiter WASP-4b. WASP-4b has a mass of 1.22~$M_\mathrm{J}$ and a planetary radius of 1.34~$R_\mathrm{J}$ \citep{2008ApJ...675L.113W,2010A&A...524A..25T}. We study WASP-4b because its equilibrium temperature is 1700~K, making it a favorable target due to a large predicted atmospheric signal. WASP-4b is also an excellent candidate for transmission spectroscopy due to its having a comparison star of near-identical type and similar magnitude $\sim 1.3$ arcmin away. 
Finallly, its G7 host star is only moderately active, with chromospheric Ca II H\&K line emission ratios of $\log (R'_\mathrm{{HK}}) = -4.865$ (see \citealt{noyes84}, \citealt{knutson10} and references therein). 

WASP-4b has been studied intensively using RV and transit techniques to refine the system parameters and attempt to constrain the stellar rotation period, which is between 20 and 40 days \citep{2008ApJ...675L.113W,2009MNRAS.399..287S,2010A&A...524A..25T,2011ApJ...733..127S,2011AJ....142..115D,2012A&A...539A.159N,2013MNRAS.434...46H}. Observations have also found that the stellar rotation axis is aligned with the planet's orbital axis \citep{2010A&A...524A..25T,2011ApJ...733..127S} and have ruled out planetary companions more massive than 2.5~R$_\mathrm{Earth}$ via transit timing variations \citep{2009AJ....137.3826W,2013ApJ...779L..23P,2013MNRAS.434...46H}. 

Despite detailed study of the WASP-4 planetary system, however, secondary eclipse observations have not constrained the composition \citep{2011ApJ...727...23B,2011A&A...530A...5C,2014ApJ...785..148R} and there are no robust transmission spectral data available for this planet. The transmission spectrum of WASP-4b has been observed with WFC3 in the near-infrared but problems with the treatment of detector non-linearity meant that a robust transmission spectrum could not be obtained \citep{2014ApJ...785..148R}. In addition, \citet{2012A&A...539A.159N} observed the transit of WASP-4b in 4 filters simultaneously, but the uncertainties on the measured transit depths were too large to draw conclusions about absorbers in the planet's atmosphere. 

Our aim in this paper is to provide a complete optical transmission spectrum for WASP-4b and to determine the presence or absence of atomic and molecular species such as Na~\textsc{i}, K~\textsc{i}, TiO and scattering features as well as to identify any other prominent absorbers in the optical transmission spectrum. Since this is the first result from our survey of hot Jupiter atmospheres with Gemini GMOS, we also present a complete pipeline as well as a detailed characterization of the GMOS instruments as tools for exoplanet characterization.

\section{Observations}
\label{sec_obs}

We observed four transits of WASP-4b in the optical at low resolution, using the Gemini South telescope located at Cerro Pachon, Chile. The observations were similar to the multi-object spectroscopy (MOS) observations pioneered by \citet{2010Natur.468..669B,bean11} and widely used (e.g. \citealt{gibson13,bean13,2014AJ....147..161S,2016ApJ...817..141S}). For each observation, we used the MOS mode of GMOS to observe time-series spectrophotometry of WASP-4 and a similar-magnitude comparison star simultaneously. Each observation covered a planetary transit and lasted approximately 5 hours. The comparison star used was 2MASS J23341836-4204509, and it is at a distance of $\sim 1.3$~arcmin from WASP-4. 

In order to avoid slit losses, our MOS mask had wide slits of 10~arcsec width for each star. The slits were 30~arcsec long in order to ensure adequate background sampling for each star. In order to make sure that the spectra of both stars had a similar wavelength coverage for each observation, we selected the position angle (PA) of the MOS mask to be as close as possible to the PA between the two stars (332.5 deg. E of N). 

We observed one transit of WASP-4b using the B600 grating, covering a wavelength range of 400-650~nm with ideal resolving power $R = 1688$. This transit was a pilot study to assess whether Gemini GMOS was suitable for a survey of exoplanet transmission spectra. We then began our survey of hot Jupiter atmospheres, in which we observed a further three transits of WASP-4b using the R150 grating. The R150 grating covers a wavelength range of 525-900~nm with ideal resolving power $R = 631$. Note that the ideal resolving powers for both gratings assume a slit width of 0.5~arcsec. In our case, due to using a wide slit, our resolution was seeing limited. Given the range of seeing measured in Table~\ref{obsstats}, our resolution is approximately $2-4\times$ lower than the ideal values depending on observation. Program numbers for each transit observation are given in Table~\ref{obsstats}.  


\begin{table*}
\centering
\begin{tabular}{ccccccccc}
\hline
\hline
Program ID & Observation & Grating & Exposure  & No. of & Duty & Seeing & Start & End \\
& Date (UT) & & Time (s) & Exposures & Cycle (\%) & (arcsec) & Airmass & Airmass \\
\hline
GS-2011B-Q-45 & 2011 Oct 10 & B600 & 300 & 59 & 90 & 0.6-2.0 & 1.19 & 1.25 \\
GS-2012B-Q-6\tablenotemark{a} & 2012 Oct 16 & R150 & 100 & 153\tablenotemark{d} & 80\tablenotemark{d} & 0.9-1.6\tablenotemark{e} & 1.05\tablenotemark{e} & 1.65\tablenotemark{e}  \\
GS-2013B-Q-44\tablenotemark{b} & 2013 Oct 11 & R150 & 50 & 219 & 76 & 0.4-0.9 & 1.03 & 1.51 \\
GS-2014B-Q-45\tablenotemark{c} & 2014 Sep 24 & R150 & 75 & 132 & 62 & 0.7-1.3 & 1.53 & 1.02 \\
\hline
\end{tabular}
\caption{Observing conditions for GMOS-South runs. $^\mathrm{a}$ Hereafter referred to as `transit~1' \\ $^\mathrm{b}$ Hereafter referred to as `transit~2'  $^\mathrm{c}$ Hereafter referred to as `transit~3'. \\ $^\mathrm{d}$These numbers include the first 14 exposures of the observation, which were later discarded (see text).
\\ $^\mathrm{e}$These numbers are for only the non-discarded exposures of transit~1 (see text).}
\label{obsstats}
\end{table*}

For each observation, we used the gratings in first order. For the R150 observations, the requested central wavelength was 620~nm and we used the OG515\_G0330 filter to block light blueward of 515~nm. The blocking filter was used to avoid contamination from light from higher orders. For the B600 observation, the requested central wavelength was 530~nm and no blocking filter was needed. 

For the B600 observation, we read out the full frame of the detector using $2 \times 2$ binning to reduce overheads. Three amplifiers were used simultaneously with gains approximately 2~$e^-$/ADU (recorded in the FITS headers). For all three R150 observations, we were further able to reduce overheads by reading out only regions of interest (ROI) on the detector rather than the whole frame. We used one ROI for each slit, covering the whole detector in the dispersion direction and approximately 40~arcseconds in the cross-dispersion direction. 

The GMOS-S detector was replaced in June 2014, during our survey program, in order to reduce the effects of fringing and improve red sensitivity\footnote{https://www.gemini.edu/sciops/instruments/gmos/imaging/\\detector-array/gmosn-array-hamamatsu?q=node/10004}. As a result, R150 transits~1 and~2 used the original detector, manufactured by e2v\footnote{http://www.e2v-us.com/}, while transit~3 used the new detector, manufactured by Hamamatsu\footnote{http://www.hamamatsu.com/us/en/index.html}. As with the B600 observation, three amplifiers were used for R150 transits~1 and~2 and we used the $2\times2$ readout mode to improve overheads. For transit~3, the new setup used 12 amplifiers simultaneously, which reduced overheads enough that we were able to use the $1\times1$ binning mode. The amplifier gains for transit~3 varied from 1.6 to 1.85~$e^-$/ADU. 

Table~\ref{obsstats} shows the observation log for each transit. Exposure times were chosen to keep count levels between 10,000 and 30,000 peak ADU and well within the linear regime of the CCDs. For R150 transit~1, guiding was temporarily lost after the 14th exposure, and a shift of 2.5~arcsec was applied once the target was re-acquired, to better center the targets on the detector. We found that the first 14 exposures showed a significantly lower flux compared to the rest of the exposures, as well as a different wavelength solution, and so we excluded these from the analysis. The pointing remained stable for the rest of the night, with no other gaps in the observation. There were no gaps in any of the observations for the other three transits.


\section{Data Reduction}
\label{sec_data_reduction}

In order to reach the high precisions required for exoplanet transmission spectroscopy, we developed a custom pipeline for reducing our GMOS data. Our pipeline was initially based on the steps described by \citet{2010Natur.468..669B,bean11} and \citet{2014AJ....147..161S} but we also include additional steps primarily focused on correcting for time- and wavelength-dependent shifts. We describe our data reduction pipeline and procedure in this section. The data reduction pipeline will be made public.

\subsection{2D Image Processing}

The initial step is to combine the images from all amplifiers per frame, multiply by the gains and subtract the bias level. Each amplifier image for each frame has an overscan region and we used these to measure the bias level. We also took a series of bias frames the day after each science observation and found no significant difference in the lightcurves produced using the bias frames as the bias measurement and the lightcurves produced using the overscan region as the bias measurement

The second step is to identify and remove cosmic rays, which we did by looking for and removing outliers in time for each pixel. For each pixel, the pipeline takes batches of 10-20 exposures at a time and replaces flux values $> 5$~$\sigma$ deviant from the batch with the median of the batch. The reason for using batches rather than the entire time series is that there can often be large flux variations in the time series due to atmospheric and instrumental conditions, inflating the standard deviation and causing cosmic rays to go undetected. Using our method, a few percent of array elements were flagged for each science observation.

The next step of the pipeline provides the option to flat-field the data. For each observation, a series of flats was taken during the day either before or after the science observation, using the same MOS mask as that used for the science observations. The pipeline median-combines each series into a master flat for each observation while also performing outlier rejection. It then fits for and removes the instrumental response function using a smooth function. 

For our WASP-4 data, however, we chose not to flat-field for several reasons. For the B600 transit, technical problems meant that we only obtained 2 flat fields, meaning that flat-fielding added noise due to low count levels overall. For R150 transits~1 and~2, both the flat-fields and science frames show fringing at the 10~\% amplitude. We found that the scatter in the transit lightcurves redward of 700~nm was $10\times$ photon noise without flat-fielding and even higher when performing flat-fielding. On inspection of the frames, we found that noise is added by flat-fielding because the phase, period and amplitude of the fringe pattern is significantly different between the flat fields and the science frames even when using the same PA and the same telescope altitude for both flats and science. For R150 transit~3, the fringe amplitude is an order of magnitude lower than in the other R150 transits, and we were able to obtain 200 flats with a median count level of $\sim 10,000$. However, flat-fielding still increased the scatter redward of 700~nm by 10-20~\%. We attribute this to low-levels of fringing still being present. Since flat-fielding did not improve scatter blueward of 700~nm, we also chose not to perform flat-fielding for transit~3. 

After optional flat-fielding, a new step is performed by our pipeline: the removal of columns of shifted charge. These artifacts occur only when using the e2v detector. In these columns, counts appear to be intact but shifted in the cross-dispersion direction. The pipeline identifies the shifted columns by searching for deviant flux outside the slit using the following procedure. For each column, the flux is summed over 20 out-of-slit pixels in the spatial direction and then compared to the equivalent sums for the neighboring 20 columns. Columns with fluxes deviant by 3~$\times$ the median absolute deviation of the neighboring 20 columns are flagged and removed from further analysis. This procedure additionally removes columns close to the edges of the individual CCDs in the detector array, which also display anomalous fluxes outside the slit, as well as any bad columns with anomalously low or high counts. The procedure identified 0.5-12~\% of columns, with R150 transit~3 having significantly fewer removed columns, likely owing to the improved performance of the new CCDs.

The pipeline also contains the option to measure and correct for spectral tilt, which was observed by \citet{2014AJ....147..161S}. In the presence of spectral tilt, the spectral lines are diagonal rather than vertical, which can complicate background removal. Spectral tilt occurs because of geometric distortion in the instrument optics, which increase away from the center of the field of view. In addition, spectral tilt can also be caused by slit tilt, in which the object slits in the mask are tilted with respect to the field of view in order to place two or more objects in a single slit. We do not, however, use tilted slits in our masks and we do not detect any spectral tilt in the WASP-4 spectra. This is likely because our two stars are closer to the center of the field of view than those observed by \citet{2014AJ....147..161S}. We found that performing a tilt correction had no effect on our final lightcurves and therefore do not perform this correction for the WASP-4 dataset.

\subsection{Spectral Extraction and Wavelength Calibration}

After the image processing detailed in the above section, we performed an optimal spectral extraction based on the algorithm described in \citet{horne86}. We initially used a range of aperture sizes and then selected the aperture that produced the lowest rms in the out-of-transit lightcurves. The aperture radius used was 25 pixels (3.65~arcsec) for the B600 observation and R150 transits~1 and~2. For R150 transit~3, the aperture radius used was 50 pixels (4.0~arcsec).

Background values were estimated by using a linear fit to the fluxes in each column in the background region. This provided the best fit to the background fluxes compared with using a median value or higher-order fits. For the R150 observations, the background level was $\sim 3 - 10$~\% of the stellar flux level, with the higher background ratios occurring at longer wavelengths. For the B600 observation, the background was 5 - 30~\% of the stellar spectral flux, with the ratio decreasing with increasing wavelength.

After spectral extraction, we performed wavelength calibration using CuAr lamp spectra taken on the same day as each science observation. To obtain the CuAr spectra at high resolution, we used a separate MOS mask to that used for science, which had the same slit positions and slit lengths as the science mask but slits of only 1~arcsec width. We used the same grating and filter setup as the corresponding science observation. 

\begin{figure}[h]
\epsscale{1.1}
\plotone{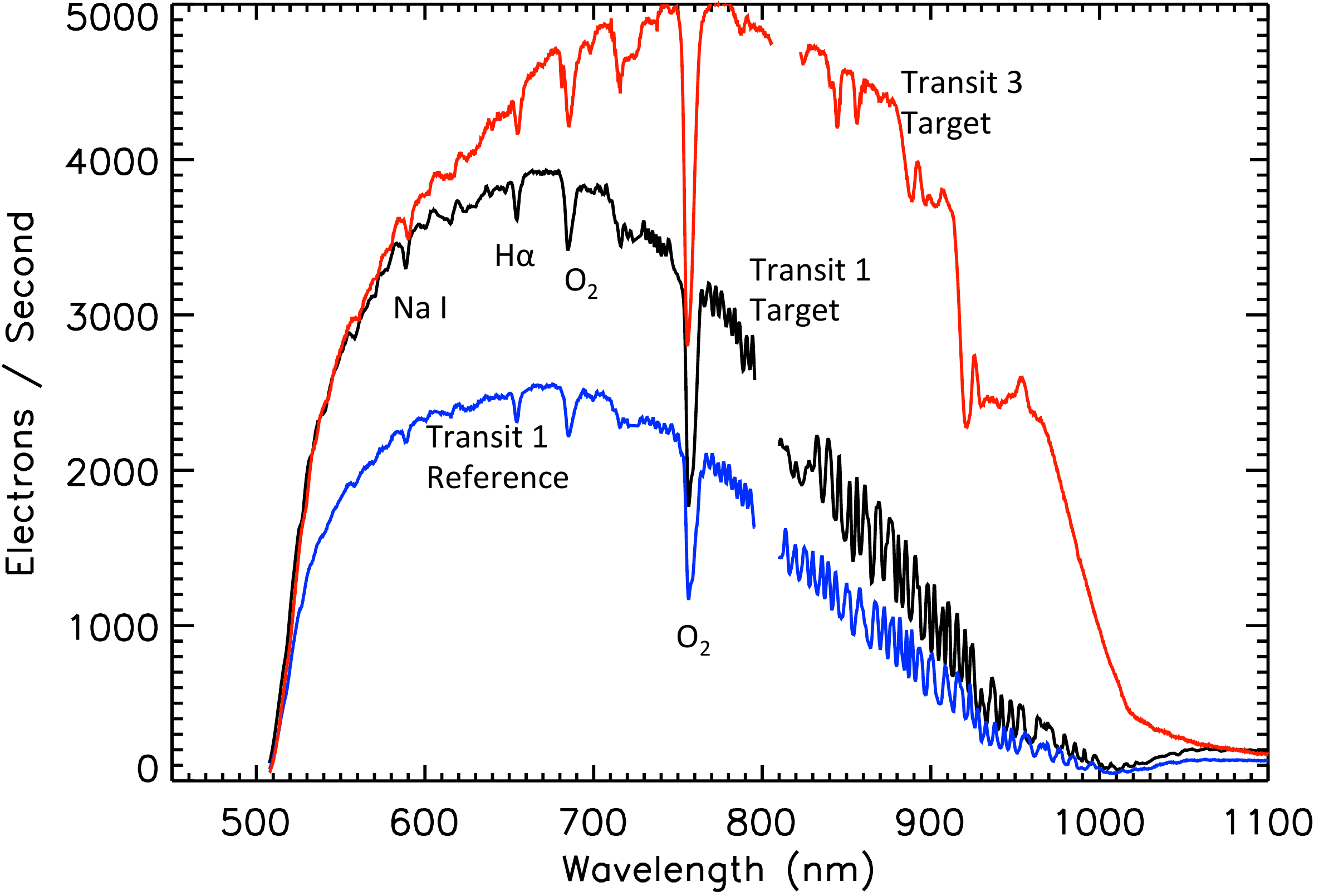}
\caption{Sample extracted spectra for R150 transit~1 as well as a sample extracted spectrum from R150 transit~3 to illustrate the improvement in sensitivity and reduction in fringing from the new detector. All spectra were extracted at a similar airmass, around 1.25, and normalized by exposure time. Since the pixels in the two detectors are different sizes, the spectrum from R150 transit~3 has a different number of pixels in the covered wavelength range than the other spectra. We therefore also normalized the spectrum from R150 transit~3 by the ratio of the pixel sizes in order to compare all the spectra on the same scale. Prominent stellar and telluric features are labelled. Fringing is apparent at wavelengths longer than 700~nm for transit~1. The gaps in wavelength coverage are due to the physical gaps between individual CCDs in the detector.}
\label{fig_r150_extractedspec}
\end{figure}

We performed a manual identification of known spectral features in the CuAr spectra using the \textsc{iraf} \textsc{identify} package, and performed a separate identification for each slit. We then performed a linear fit to the pairs of wavelength solution vs. pixel number to obtain an initial wavelength solution. The wavelength solution was then refined by comparison with known stellar and telluric lines. Example wavelength-calibrated extracted spectra are shown in Figure~\ref{fig_r150_extractedspec} for R150 transit~1. The final uncertainties in the wavelength solution are approximately 1~nm for all observations, which is only 10~\% of the bin widths used in the final transmission spectrum. 

\subsection{Instrumental Corrections Using 1D Spectra} 
\label{sec_1d_corrections}

Before producing the final lightcurves, we performed further reduction of the extracted 1D spectra. The main reason for additional reduction is that the stellar spectra shift in the dispersion direction during the course of an observation, with the amplitude of the shifts being a function of time, wavelength and position in the focal plane. The result is that the spectra `stretch' over time, meaning that a given pixel will not sample the same wavelength in each exposure. This effect therefore needs to be accounted for in order to build transit lightcurves that sample a constant wavelength region over time. Failure to account for this effect can introduce spurious slopes into transmission spectra (discussed in Section~\ref{sec_noise_discussion}).

We first investigated possible causes of the stretch in order to develop a physically-motivated model with which to correct it. Since GMOS is not equipped with an atmospheric dispersion compensator (ADC), the first effect to model is that of differential atmospheric refraction. We used the \textsc{idl} code \textsc{diff\_atm\_refr} written by Enrico Marchetti\footnote{http://www.eso.org/gen-fac/pubs/astclim/lasilla/diffrefr.html} and based on \citet{1982PASP...94..715F} and \citet{1999A&AS..136..189A} to compute differential atmospheric refraction in arcseconds relative to 500~nm for the wavelengths and airmasses covered in our observations. We used the average atmospheric conditions for Cerro Pachon measured by Gemini throughout multiple observing seasons, which are $T=7^o$~C, $RH=14.5$~\%, $P=836$~mbar\footnote{https://www.gemini.edu/sciops/instruments/gmos/itc-sensitivity-and-overheads/atmospheric-differential-refraction}. Cerro Pachon is at latitude -30.24 degrees.

\begin{figure}[h]
\epsscale{1.0}
\plotone{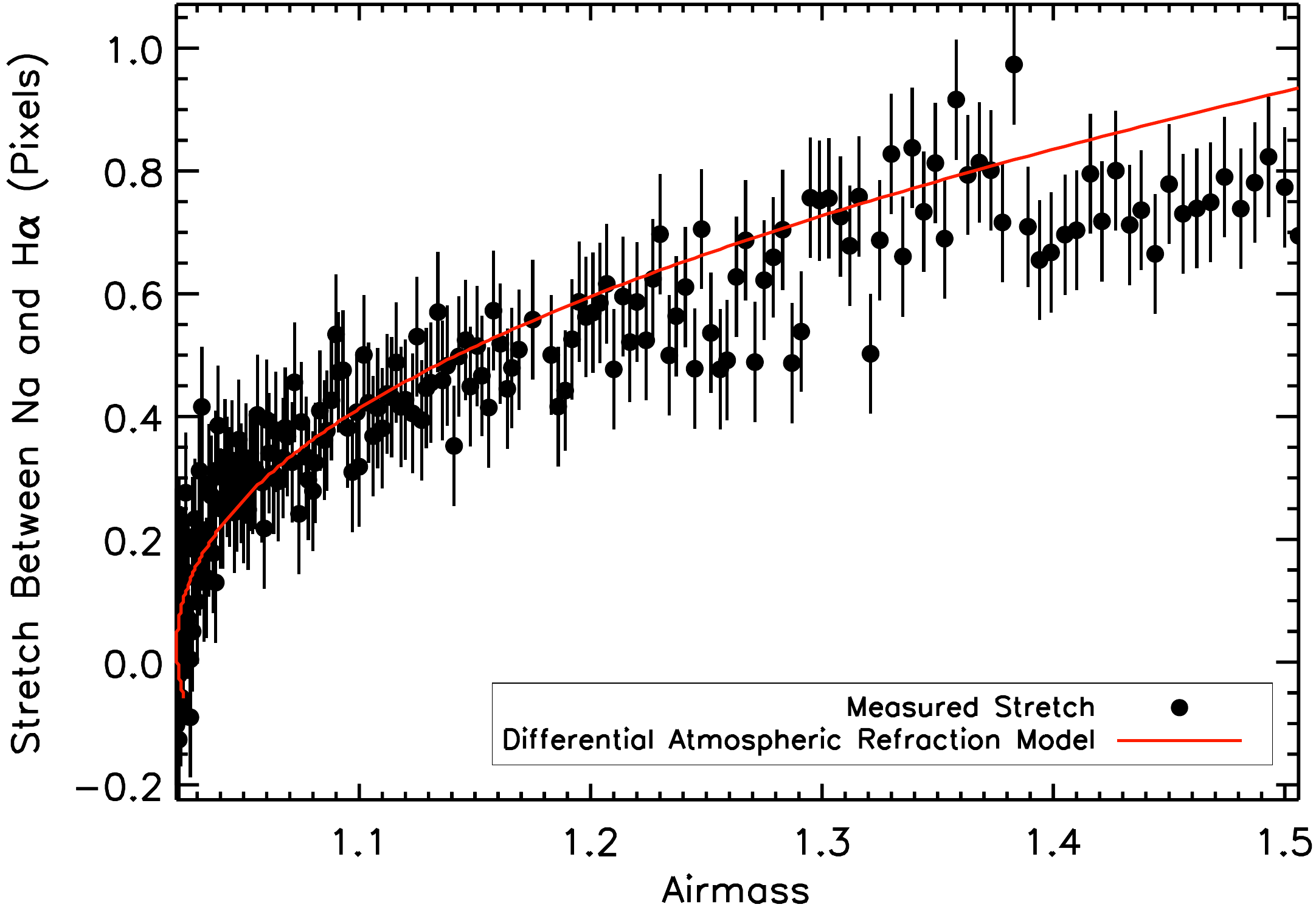}
\caption{Dispersion direction shift measured in the Na~\textsc{i} (589.3~nm) core minus dispersion direction shift measured in the H$\alpha$ core (656.3~nm) vs. airmass for R150 transit~2. The shift values for each feature were determined by cross-correlating the spectra using a small 100-pixel range around each feature. Overplotted is the expectation from differential atmospheric refraction. The reduced $\chi^2$ for the model fit is~1.}
\label{fig_daf_data}
\end{figure}

We then converted the displacement at each wavelength and at each time (given the telescope PA) from arcseconds into pixels in the dispersion direction using the pixel scales of 0.073 arcsec/pixel for the e2v detector and 0.080 arcsec/pixel for the Hamamatsu detector\footnote{https://www.gemini.edu/sciops/instruments/gmos/\\imaging/detector-array}. Figure~\ref{fig_daf_data} shows the differential atmospheric refraction model along with the `stretch' in pixels between the Na~\textsc{i} feature core (589.3~nm) and H$\alpha$ feature core (656.3~nm) for R150 transit~2.
The uncertainties on the data points are the cross-correlation uncertainties, which we estimate using the standard deviation of the measured lags after removal of a linear function of lag vs. time. The uncertainties are $0.1-0.2$~pixels per exposure. 

While the broad trend is the correct order of magnitude, the model does not fit the data well in all places. Furthermore, extensive evaluation of our Survey dataset of 9 planets shows that the differential atmospheric refraction model is often a poor description of the observed stretch (comparing the model with the data gives reduced $\chi^2 \sim 5$ on average across 29 transit observations in total). We found that varying the site parameters within reasonable ranges had a negligible effect on the model fit. This suggests that there is an additional effect not explained by differential atmospheric refraction. 

We found that the stretch is not seen in the background spectra, only in the stellar spectra, which indicates that focal plane variations such as flexure are not responsible for the additional effect. We therefore assume that it is an optical variation, possibly depending on the location of the star in the slit along the spectral axis. Since we are unable to access guider camera images from our observations, we were unable to make a model of the possible effect. 

We therefore produced an empirical solution that can be used in cases where the differential atmospheric refraction model is a poor fit to the observed spectral stretch. In the empirical method, we selected several prominent spectral features in the stellar spectrum by hand in pixel space. We then produced multiple time-series spectra, with each time series cross correlated and interpolated using a different spectral feature. When constructing the lightcurves for each spectral bin, we use the time series cross-correlated to the closest spectral feature in wavelength to that bin. 

We tested the empirical model by using the case of WASP-4, in which we know that the differential atmospheric refraction model is a good fit to the observed spectral stretches. We found that the transmission spectrum of WASP-4b produced using the empirical method is insignificantly different to that produced using the differential atmospheric refraction model, indicating that the empirical method is also a good fit to the observed spectral stretches. Since the empirical method uses the data themselves to form a model, however, we are confident that it will be equally effective in cases where the differential atmospheric refraction model is inadequate. We use the empirical model to produce the final transmission spectrum of WASP-4b.

The final step in the reduction process, after accounting for time-dependent shifts, is to correct for the fixed dispersion-direction offset between our two stars. This offset is constant in time and occurs because the PA of the instrument is not exactly the same as the PA between the target and reference star. We used cross-correlation to measure the offset, which was between 5 and 8.5 pixels for our 4 observations. We then interpolated the reference star's spectra onto the target star's wavelength grid, being careful to omit bad columns for both spectra (which are the same columns on the detector but are at different wavelengths for each star). 

\section{Analysis}
\label{sec_analysis}

\subsection{Transit Parameters}
\label{sec_whitelight}

Our first goal is to measure the system parameters for WASP-4, which are: orbital inclination, $i$, the system scale, $a/R_\star$, central transit time, $T_0$, orbital period, $P$, radius contrast, $R_\mathrm{P}/R_\star$, for each bandpass and linear limb-darkening coefficients, $u$, for each bandpass. To begin with, we fitted each transit separately for system parameters, in order to test whether all transits produce consistent results. This also allows us to use different wavelength ranges for the different R150 transits depending on the effects of fringing.  

For these measurements, we construct and fit `white' lightcurves for each transit, which are summed spectrally and hence provide the highest signal-to-noise measurements possible. We used the following procedure to produce the white light curves for each transit. First, we summed the flux spectrally over all wavelengths for each exposure for WASP-4 to produce a lightcurve for the host star. We then repeated the process for the reference star, summing over the same wavelength region for both stars. We then divided each lightcurve of WASP-4 by the corresponding comparison star's light curve in order to reduce the effects of instrumental and Earth-atmospheric variations during the observation. Due to the high amplitude of fringing redward of 700~nm in R150 transits~1 and~2, we excluded wavelengths redder than 700~nm from these white lightcurves in order to reduce scatter. 

To measure the transit parameters, we fitted each white lightcurve with a model that contained the transit plus a parameterization of the systematics. To parameterize the systematics, we used a linear model of planet orbital phase $\times$ baseline stellar flux for the B600 observation and R150 transits~1 and~3. For R150 transit~2, we also included a linear function of airmass in the model. The full range of possible parameterizations for the systematics that we investigated are detailed in Section~\ref{sec_parameterization}. We used the \textsc{batman} package to model the transit \citep{2015PASP..127.1161K} using a linear limb-darkening prescription. 

We used the Python MCMC package \textsc{emcee} to fit the parameters describing the transit and the systematics simultaneously \citep{2013PASP..125..306F}. We fitted each observation separately. We used the routines of \citet{2010PASP..122..935E} to convert the calendar dates in the headers of each frame of each observation to BJD$_{\mathrm{TDB}}$.

\begin{figure*}
\epsscale{1.0}
\plotone{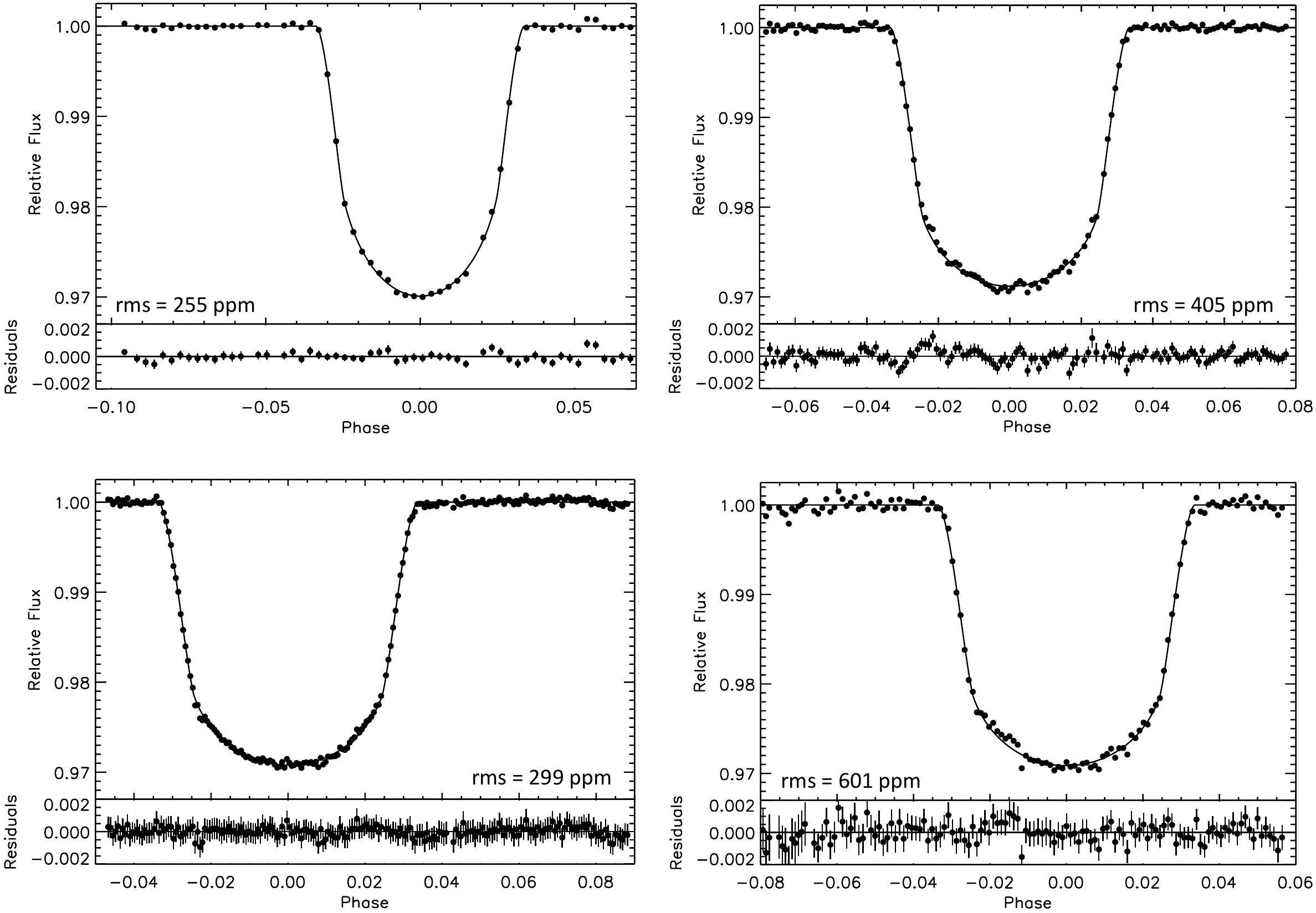}
\caption{GMOS white light curves for the B600 observation (top left), R150 observation~1 (top right), R150 observation~2 (bottom left) and R150 observation~3 (bottom right). Each plot shows the target's lightcurve divided by that of the reference star after correcting for systematics. In each case, the best-fitting transit model is shown. Error bars on the lightcurves are the initial uncertainties from photon noise. Error bars in the residuals are re-scaled after the fit using the $f$ parameter (see text).}
\label{whitelight_corrected}
\end{figure*}

Figure~\ref{whitelight_corrected} shows the white lightcurves for each observation after removing the best-fitting models describing the systematics. Also overplotted are transit lightcurve models constructed using the best-fitting system parameters and the analytical transit models of \citet{mandelagol02}. The best-fitting system parameters are given in Table~\ref{system_params_table}. 

We ran each fit with 100 simultaneous chains, testing for convergence of the chains using the Gelman-Rubin criterion (see \citealt{ford06} and references therein). We found that 10,000 steps was sufficient for the chains to converge and that additional steps did not change the results. We trimmed the first 10~\% of the steps as the burn-in and then took the median value of each parameter's distribution as the parameter value. The uncertainties given on each parameter are the 68~\% confidence intervals. Our posterior distributions are well-behaved and the parameter values with peak probability are equal to the median values.

Our initial uncertainties are derived from the optimal extraction routine, assuming photon noise and read noise only. However, \textsc{emcee} allows us to include a parameter, $f$, in the likelihood function which is the additional uncertainty compared to photon noise, such that the likelihood function becomes

\begin{equation}
\ln(L) = - \frac{1}{2} \sum \left[ \frac{(y-m)^2}{s^2} + \ln{(2 \pi s^2)} \right] ,
\end{equation}
where $L$ is the likelihood, $s^2 = \sigma^2 + f^2 m^2$, $\sigma$ is the predicted (photon noise) uncertainty, $y$ are the data and $m$ is the parameterized model. The parameter $f$ is therefore included in our uncertainties on the final measured transit parameters. We report how far from photon noise we are in each observation in Table~\ref{system_params_table}. 

We used flat priors for all parameters except the limb-darkening coefficients, for which we used a Gaussian prior around the values obtained from Kurucz (1993) stellar atmosphere models (see Section~\ref{sec_parameterization}). In order to take into account uncertainties in the model and in measured values, we chose the 1-$\sigma$ width of the Gaussian to be 0.2, which is double the largest uncertainty observed in the literature \citep{2013MNRAS.434...46H}. We found that fitted parameters did not significantly change when using a flat prior and that including priors on the other fitted parameters did not significantly affect the results. While the $R_P/R_\star$ values are consistent across all three R150 observations, we note that the fitted values of $i$ and $a/R_\star$ are not consistent between the fits to the different transits. These differences are discussed in Section~\ref{sec_syspar} but do not affect our final transmission spectrum significantly. 
 
\begin{table*}
\centering
\begin{tabular}{ccccccc}
\hline
\hline
Observation & $i$ & $a/R_\star$ & $T_0$ (BJD$_\mathrm{TDB}$) & $R_\mathrm{P}/R_\star$ & $u$ & Noise/Photon \\
\hline
B600 Observation & $87.9_{-0.9}^{+2.8}$ & $5.37_{-0.09}^{+0.08}$ & $2455844.66287087 \pm 9 \times 10^{-5}$ & $0.1536 _{-0.0012}^{+0.0014}$ & 0.72$\pm 0.021$ & 2.6 \\
R150 Observation~1 & $86.5_{-0.29}^{+0.28}$ & $5.28_{-0.04}^{+0.04}$ & $2456216.69122887  \pm 6 \times 10^{-5}$ & $0.1564 \pm 0.0006$ & 0.57$\pm 0.017$ & 2.3 \\
R150 Observation~2 & $87.8_{-0.41}^{+0.34}$ & $5.40_{-0.02}^{+0.03}$ & $2456576.67555887  \pm 5 \times 10^{-5}$ & $0.1567 \pm 0.0005$ & 0.52$ \pm 0.009$ & 1.6 \\
R150 Observation~3 & $89.7_{-1.8}^{+1.8}$ & $5.50_{-0.05}^{+0.03}$ & $2456924.61561187 \pm 6 \times 10^{-5}$ & $0.1551 \pm 0.0006 $ & 0.48$ \pm 0.017$ & 1.2 \\ 
\hline
\end{tabular}
\caption{Planetary system parameters for the WASP-4 system. Note that R150 observations~1 and~2 only include wavelengths blueward of 700~nm while observation~3 covers the whole R150 bandpass. The measured $R_\mathrm{P}/R_\star$ for R150 observation~3 using only wavelengths blueward of 700~nm is consistent with the others at $0.1566^{+0.0007}_{- 0.0007}$. The final column shows how many times photon noise our uncertainties are.}
\label{system_params_table}
\end{table*} 
 
After performing separate fits to each transit, we then attempted to improve our estimate of the parameters by fitting all 4 transits simultaneously with single values of $i$, $a/R_\star$, $P$, $T_0$ and with one transit radius and limb-darkening coefficient for each band. However, the fit was complicated by slight non-linearities in the ephemeris as a function of time. We therefore do not provide the results in this section. The results are provided, along with a discussion, in Section~\ref{sec_syspar}. 
 
\subsubsection{Model Prameterization}
\label{sec_parameterization}

In order to ensure that our model describing the systematics was accurate, we investigated a variety of parameterizations, from which we selected our model: a linear function of phase, linear and quadratic functions of airmass, mean sky brightness, mean FWHM of the 2D spectra, $x$ and $y$ shifts of the spectrum on the detector and any variable parameters in the FITS headers. While we do not see an obvious trend associated with the cassegrain rotator position angle (CRPA) as seen by \citet{2014AJ....147..161S} for WASP-12b, we also tested models including functions of CRPA. 

For each parameterization, we modeled the transit using the analytic models of \citet{mandelagol02} and then fitted the transit + systematics model to the white lightcurves simultaneously. The fits were performed using the \textsc{idl} package \textsc{mpfit}, which performs Levenberg-Marquardt least-squares fitting \citep{markwardt09}. For each observation, we selected the model with the lowest value of the  Bayesian Information Criterion (BIC), which is given by $\mathrm{BIC}=\chi^2+k\ln n$ \citep{schwarz78}, where, $k$ is the number of free parameters and $n$ is the number of data points. The selected models were then used in the MCMC fits.

In addition, we also investigated whether using higher-order limb-darkening prescriptions improved the transit lightcurve fits. We obtained similar results and uncertainties for the remaining system parameters and so we elected to use a linear limb-darkening prescription. This is because the simpler limb-darkening law allows us to fit for the coefficient using the transit lightcurve itself, while using a higher-order prescription requires the coefficients to be fixed due to the complexity of the model. This is, however, only true for the white lightcurves. For the wavelength-dependent lightcurves constructed using smaller wavelength bins, our precision is lower and is insufficient to fit the limb-darkening coefficient (see below).

\subsection{Wavelength-Dependent Lightcurves}
\label{sec_binned_LC}

\subsubsection{Absolute Wavelength-Dependent Transit Depths}

In order to produce the transmission spectrum, we need to measure the transit depth as a function of wavelength. We therefore constructed lightcurves in multiple spectral bins so that we can fit each wavelength bin with a transit depth. The transit lightcurves were constructed similarly to the white lightcurve but in each case we summed the observed spectrophotometric time series over custom wavelength bins rather than over all wavelengths. 

In order to have the most precise measurement of the transmission spectrum, we selected wavelength bin sizes which minimized the out-of-transit rms in the resultant lightcurves compared to photon noise. The optimum bin sizes are approximately 10-20~nm in width. In addition, we ensured that the bins closest to the potential alkali features were centered exactly on the cores of those features, where signal is largest, in order to maximize the possibility of detecting the alkalis in the planetary atmosphere. 

As with the white lightcurves, we used the \textsc{emcee} package to fit a transit+systematics model simultaneously to each binned lightcurve. Since only the transit depth and limb-darkening depend on wavelength, we fixed all other system parameters to the best-fitting values from the white lightcurves. For the R150 transits, we used the weighted mean of the system parameters for all R150 transits as the best-fitting values. Parameters governing the systematics were allowed to vary. 

Unlike our method for the white lightcurves, we found that the lowest BIC was produced by using the non-linear limb-darkening law given in \citet{sing10}, with the coefficients fixed to the values derived from the Kurucz (1993) models. To obtain the limb-darkening coefficients, we used a stellar atmosphere model with $T_\mathrm{eff}=5500$~K, $\log g[\mathrm{cgs}]=4.5$ and [M/H] = 0.0 \citep{2008ApJ...675L.113W}

When performing these fits, we investigated the effect on the measured transmission spectra of adjusting the system parameters within their uncertainties. We noticed that the variation in $R_P/R_\star$ as a function of wavelength, $\Delta R_\mathrm{P}/R_\star$, agreed well between the different system parameters used but that the absolute $R_P/R_\star$ values did not at the 2-$\sigma$ level. This is not surprising since we find differences in the fitted parameters between the individual visits (Table~\ref{system_params_table}). We therefore decide to limit ourselves to measurements of  $\Delta R_P/R_\star$ as a function of wavelength for the rest of the analysis. Such a measurement is more robust than the absolute measurements and is all we require for the measurement of the transmission spectrum. 

\subsubsection{Common Mode Corrections}
\label{sec_common_mode}

Since we are limiting ourselves to measurements of $\Delta R_P/R_\star$, we no longer need to preserve information about the absolute transit depths in our lightcurve fits. This means that we can employ common-mode corrections, which can provide a superior removal of systematic trends at the expense of losing information about absolute transit depth. 

Common-mode corrections are described in detail by \citet{2014AJ....147..161S}, who employ these corrections for GMOS spectrophotometry. We follow the same procedure for our analysis. In the common-mode correction, we assume that the majority of trends have the same time-dependence in all wavelengths. These wavelength-independent trends can then be removed from the wavelength-dependent lightcurves by subtracting the trends measured in the white lightcurve. This procedure therefore does not require a parametric model for the wavelength-independent systematics and can therefore model trends that we are unable to model parametrically.

To perform the common-mode correction, we first fitted the white light curve for a transit + baseline stellar flux level + systematic model and removed the transit model only, leaving residuals which now contain the systematics. The residuals characterize the systematic trends. We then normalized these residuals by the best-fitted baseline stellar flux level and divided each wavelength-dependent lightcurve by these before fitting for the transit and a linear function of time. The fit of a linear function of time is to account for low-level wavelength-independent trends not removed by the common-mode correction.

Compared with the absolute transit depth fits, the uncertainties on the fitted $\Delta R_P/R_\star$ decreased by 10~\% blueward of 700~nm and decreased by 20~\% redward of 700~nm when using the common-mode correction. Table~\ref{photnoise} shows how far we are from photon noise in each of our bins. The values of $\Delta R_P/R_\star$ themselves were well within 1-$\sigma$ of the values determined using the absolute method. We therefore used the common-mode subtracted lightcurves to produce the final transmission spectrum. We additionally note that the same results were obtained by normalizing and then subtracting the white lightcurve itself from each normalized spectral lightcurve and fitting for the differential transit depth and limb-darkening only.

Figures~\ref{fig_b600_spectralLC},~\ref{fig_r150_spectralLC1} and~\ref{fig_r150_spectralLC2} show all the transit lightcurves in each of the spectral bins after common-mode subtraction along with the corresponding residuals after fitting for the transit and linear function of time. For the B600 data, we found that several lightcurves displayed much higher scatter than neighboring lightcurves. These were found to be close to bad columns and so were also removed from the analysis. However, they are still plotted in Figure~\ref{fig_b600_spectralLC}.

\begin{table*}
\centering
\begin{tabular}{cc}
\hline
\hline
Wavelength & RMS / \\
(nm) & Photon Noise \\
\hline
 440.2   - 450.6  &   1.00     \\
 450.7   - 460.5  &   2.87     \\
 460.2   - 468.9  &   1.47     \\
 469.1   - 477.9  &   1.71     \\
 487.0   - 495.8  &   2.28     \\
 496.0   - 504.7  &   1.35     \\
 504.9   - 515.1  &   1.58     \\
 515.2   - 524.1  &   2.88     \\
 524.8   - 534.6  &   2.26     \\
 534.8   - 544.6  &   1.76     \\
 564.8   - 574.7  &   1.37     \\
 574.8   - 585.2  &   3.26     \\
 585.3   - 595.3  &   2.41     \\
 594.9   - 603.6  &   2.39     \\
 612.8   - 621.5  &   3.40     \\
 621.7   - 630.5  &   2.33     \\
 \hline
 \end{tabular}
\begin{tabular}{cccc}
\hline
\hline
Wavelength (nm) & & RMS / Photon Noise \\
& Obs 1 & Obs 2 & Obs 3 \\
\hline
 573.6  -  583.2  &   1.69  &   2.31  &   1.54   \\
 583.6  -  593.3  &   2.44  &   1.61  &   1.53   \\
 594.3  -  605.1  &   2.76  &   1.79  &   1.44   \\
 605.4  -  616.1  &   1.96  &   1.82  &   1.76   \\
 616.5  -  627.2  &   1.17  &   1.62  &   1.96   \\
 627.6  -  638.3  &   1.89  &   3.76  &   1.41   \\
 638.7  -  648.7  &   2.29  &   2.52  &   2.09   \\
 660.1  -  669.5  &   2.66  &   2.13  &   2.14   \\
 670.2  -  679.5  &   2.21  &   1.60  &   1.46   \\
 679.9  -  689.9  &   2.58  &   2.64  &   2.17   \\
 690.3  -  700.3  &   2.32  &   1.53  &   3.59   \\
 722.1   - 746.4    & - & - & 3.12 \\
 801.7   - 824.9    & - & - & 2.36 \\
 823.5   - 846.6    & - & - & 2.03 \\
 845.2   - 868.4    & - & - & 2.56 \\
 867.0   - 890.1    & - & - & 1.69 \\
 888.7   - 911.9    & - & - & 2.06 \\
 910.4   - 934.6    & - & - & 2.15 \\
 \hline
 \end{tabular}
 \caption{Actual RMS of lightcurves compared to the prediction from photon noise. Information for the B600 transit is on the left while information from the R150 transits are on the right.}
 \label{photnoise}
 \end{table*}
 
\subsubsection{Spectral Region Around the Sodium Doublet}
\label{Na_analysis}

We additionally searched for atmospheric absorption in the core of the Na~\textsc{i} feature at the highest resolution possible for our data. The reason for this is that the absorption cross-section is highest in the core of the feature and hence a very narrow bin could produce the most robust signal of Na~\textsc{i} if it is present in the atmosphere of WASP-4b. This is particularly important if high-altitude clouds are present in the atmosphere, which could cause only the line core to be visible and hence the signal to be washed out in our 10~nm wide bins. 

For our instrumental setup, the smallest bin size possible was 2~nm, which is limited by the resolution of the instrument in the presence of seeing. We constructed a lightcurve of this bin size centered on the Na~\textsc{i} core as well as a lightcurve in the continuum. We selected a larger, 10~nm-width, bin for the continuum to reduce noise and used the 675-685~nm region for the continuum because it is where clear-atmosphere models predict the atmospheric opacity to be lowest \citep{fortney08}. It should therefore produce the highest contrast with the opacity in the Na~\textsc{i} line core.

We additionally produced differential lightcurves in which we widened the bin centered around the Na~\textsc{i} core, similar to the method of \citet{charbonneau02}. This method cannot provide information about the line shape, but it provides the most robust possibility of finding evidence for the presence of Na~\textsc{i} since we can test whether our measured differential depths decrease with bin width as would be expected from a real signal in the doublet core.

We fitted each differential lightcurve with a differential transit depth, limb darkening and a linear function of planet orbital phase to model systematics. The results are discussed in Section~\ref{sec_discussion}.

\subsection{Transmission Spectra} 
\label{sec_trans_spec}

\begin{figure}[h]
\epsscale{1.1}
\plotone{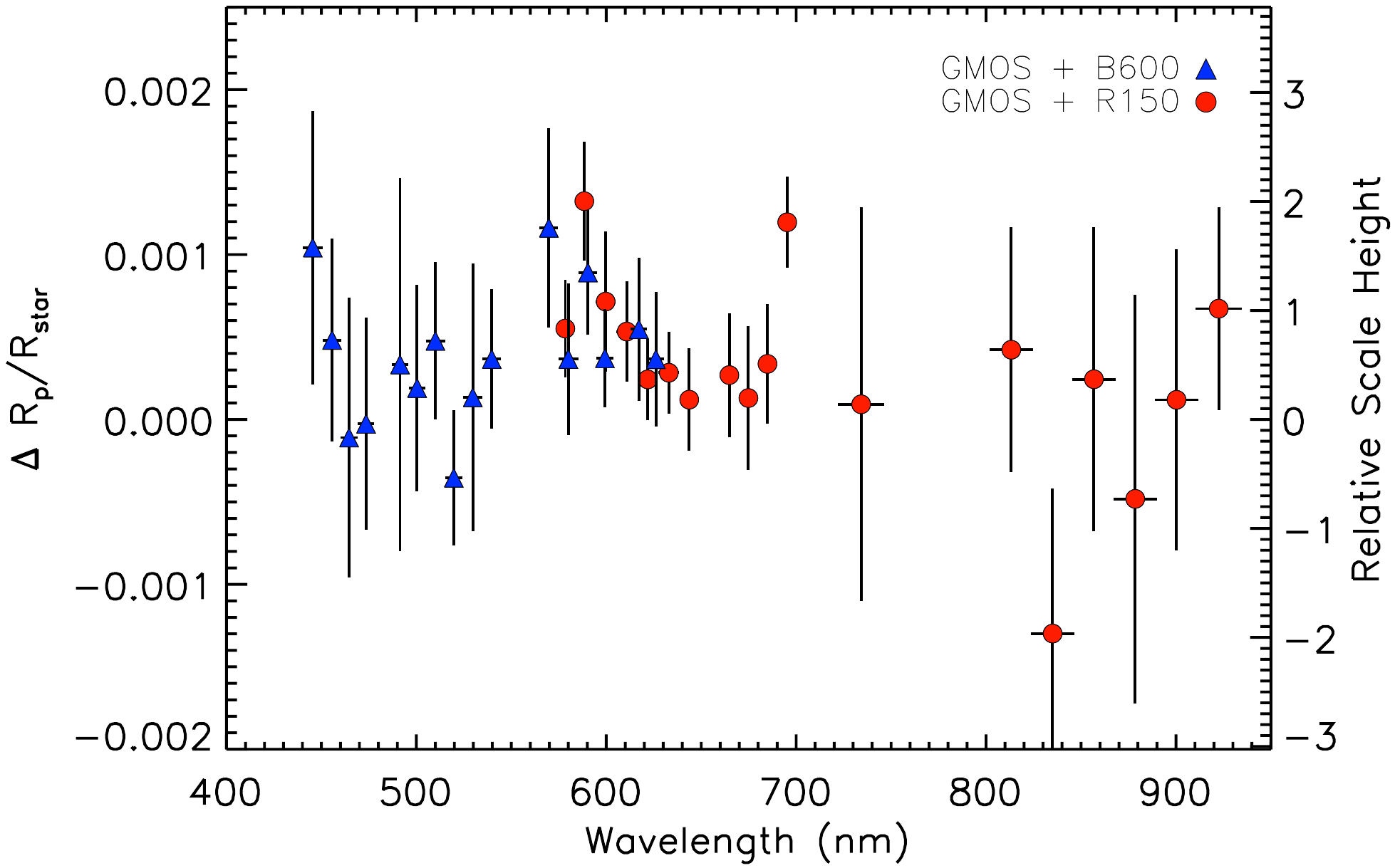}
\caption{Relative transmission spectrum from the B600 and R150 observations from the analysis in which we employed a common-mode correction of systematics (Section~\ref{sec_common_mode}). In the case of the R150 observations blueward of 700~nm, the spectrum is the weighted mean of the spectra from the individual observations. Redward of 700~nm, we were only able to use observation~3, which is why the precision is low. Points which are contaminated with bad columns or with strong telluric or stellar features have been removed (see Section~\ref{sec_noise}).}
\label{finalbroadspec}
\end{figure}

Figure~\ref{finalbroadspec} shows the transmission spectrum of WASP-4b obtained from the lightcurve fits in Section~\ref{sec_common_mode}. Since we do not have absolute transit depth information, we set $\Delta R_\mathrm{P}/R_\star = 0 $ arbitrarily at the median value of all points. In the 570-700~nm range, we have data from three R150 transits and so the $\Delta R_\mathrm{P}/R_\star$ values shown in Figure~\ref{finalbroadspec} in this wavelength range are the weighted mean from all R150 observations. All $\Delta R_\mathrm{P}/R_\star$ values shown in Figure~\ref{finalbroadspec} are tabulated in Table~\ref{table_final_transspec}. Note that we had to omit some bins due to bad columns and some due to telluric contamination (see Section~\ref{sec_noise}).

\begin{figure}[h]
\epsscale{1.15}
\plotone{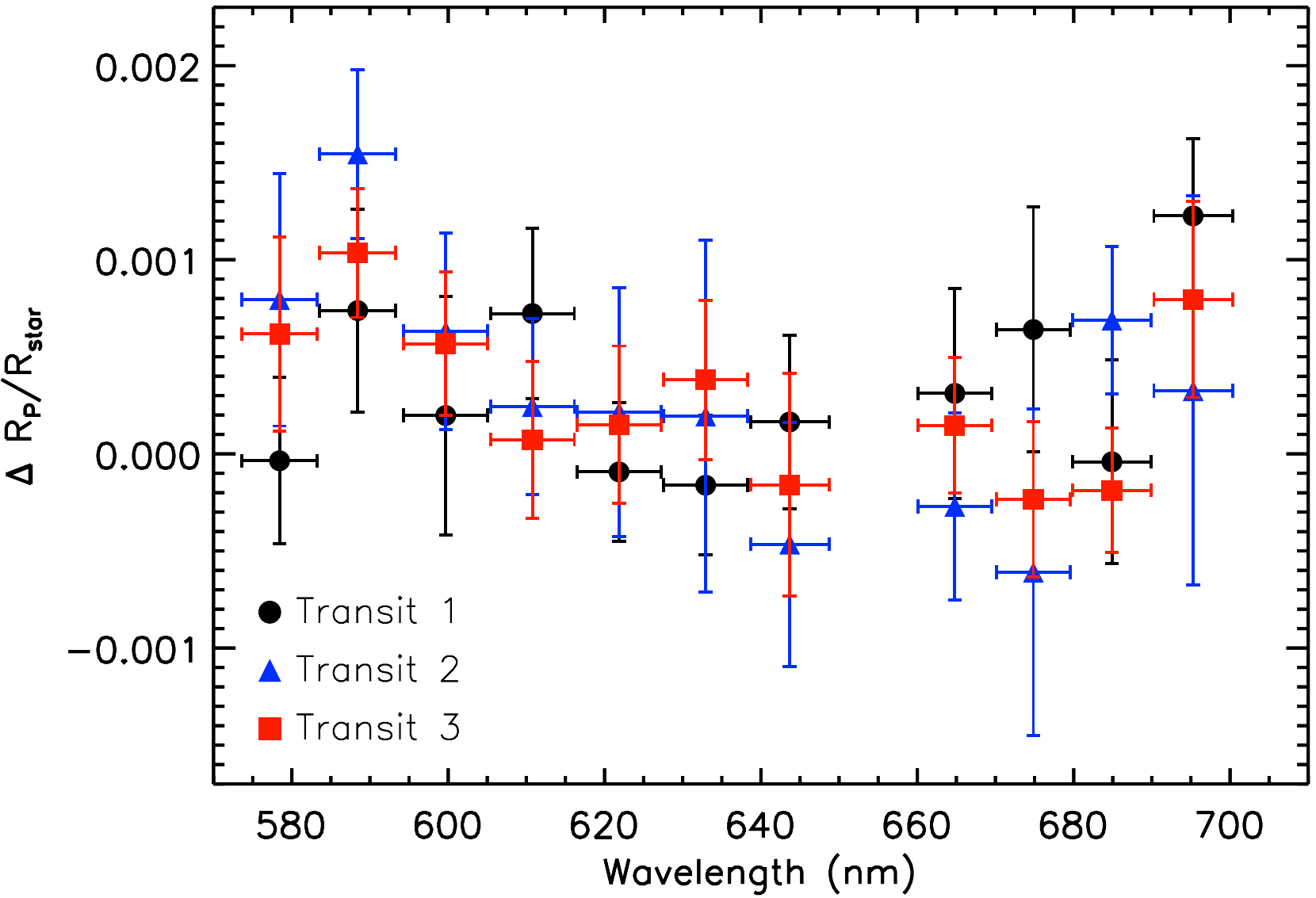}   
\caption{Relative transmission spectra obtained with the R150 grating between 570-700~nm. The spectrum from R150 observation~1 is shown with black circles, R150 observation~2 is shown with blue triangles and R150 observation~3 is shown with red squares. Despite being taken at multiple epochs, over 3 years, the transmission spectra are consistent with one another at the 1-$\sigma$ level. One bin was removed due to being within a strong  contaminating line (see Section~\ref{sec_discussion}).}
\label{r150spec}
\end{figure}

To test the reliability and repeatibility of our measurements from epoch to epoch, we checked that the $\Delta R_\mathrm{P}/R_\star$ measurements from each of the R150 transits were consistent with one another within the 570-700~nm range. Figure~\ref{r150spec} shows the transmission spectra obtained individually from each R150 transit in the 570-700~nm range, showing very good agreement between the three measurements. Table~\ref{table_r150spec} gives the corresponding measured $\Delta R_\mathrm{P}/R_\star$ values.

\section{Characterization of Uncertainties in GMOS Transmission Spectra}

\subsection{GMOS as an Instrument for Exoplanet Atmosphere Characterization}
\label{sec_noise_discussion}

The transmission spectrum of WASP-4b is the first result from a survey of~9 hot Jupiters observed in a homogeneous way with Gemini GMOS. During the course of our analysis, we used the whole survey dataset to identify recurring sources of noise in the data in order to physically model the trends seen in the WASP-4 data. Since this is the first exoplanet atmosphere survey performed with GMOS, we give a summary of the main systematic effects of the instrument in Table~\ref{gmos_limitations_table}, which will be useful for later work.

\begin{table*}
\centering
\begin{tabular}{lll}
\hline
\hline
Problem & Effect & Solution \\
\hline
Spectral stretching & Spurious Signals in & Differential atmospheric refraction correction \\
& transmission spectra & Cross-correlate spectra locally in wavelength \\
Severe Fringing (e2v) & Lightcurve noise & Cut off wavelengths $> 700$~nm \\
& 10x photon & Fixed with new detector \\
Flux discretization & Possibility of increased & No obvious cause or solution\\
& scatter in lightcurves & Likely electronic (not optical or numerical) \\
\hline
\end{tabular}
\caption{Summary of the characteristics of Gemini/GMOS when used as an instrument for exoplanet atmosphere characterization.}
\label{gmos_limitations_table}
\end{table*}

All of the effects mentioned in Table~\ref{gmos_limitations_table} have been described in earlier sections apart from a discretization of flux. This can be seen in the wavelength-dependent lightcurves in Appendix~\ref{sec_trans_lc} as periods of time where there are two envelopes of points with one set of points high and one set of points low and no points in between, for example in Figure \ref{fig_r150_spectralLC1}, top right panel, around phase $-0.06$ in the 4th bluest band. It is also visible in all of our survey data and in the published lightcurves of WASP-12b \citep{2014AJ....147..161S}. We note that this discretization is different from the ``odd-even'' effect noted in \citet{2014AJ....147..161S} for HAT-P-7b, which is caused by unequal travel time of the shutter blades for odd and even exposures, as it does not occur in every exposure and is of a higher amplitude.

The cause of this discretization is uncertain, but at our 10~nm wavelength bins other effects dominate the noise budget. At smaller bins or for brighter targets, however, the effect may become more important. We therefore provide a brief investigation into possible causes of the effect here for the purpose of guiding future work. 

First, we ruled out optical effects by noticing that the amplitude of discretization is at the photon noise level, and hence would be randomized if it were optical. We also ruled out any effect introduced by our cross-correlation procedure by constructing lightcurves without cross-correlation and observing that the effect was still present. We also extracted spectra using both \textsc{python} and \textsc{idl}, with the results being identical. The effect is therefore unlikely to be numerical. 

We conclude that the effect is likely electronic. In order to produce the amplitude of effect that we see, the discretization must occur at the level of 10 electrons/pixel, but we were unable to find a process that could be responsible. Variations in the bias overscan region are only at the~1 electron/pixel level over the duration of our observations. Additionally, the gains of the amplifiers are low enough that digitization noise is negligible. We are well within the linear range of the detectors, since our count levels do not exceed 25,000~electrons/pixel. Since we are unsure of the origin of the effect, further investigation will therefore be needed if it becomes significant in future observations.

As a final step in order to guide future GMOS and MOS observations, we conducted a test in order to quantify the effect of the observed wavelength- and time-dependent stretching discussed in Section~\ref{sec_1d_corrections}. We aim to quantify the effect in the case where we do not perform our customized correction. Since the effect is not purely due to the lack of an ADC, it could also be important for other ground-based studies, which so far have relied on simple lags as a function of time for dispersion-direction corrections.

We used the following procedure to quantify the effect on the transmission spectrum of an un-corrected time-dependent stretch. We took a single reference star spectrum and duplicated it in time to produce a time series of spectra, which we call the `target star'. We then duplicated that time series again to create a second time series, a `reference star'. The real measured shifts from each star were then applied to each time series. Simulated lightcurves were then constructed by dividing one time series by the other as we would when constructing a transit lightcurve from a target and reference star. Into this differential lightcurve, we injected a transit of constant depth as a function of wavelength and we then fitted for the transit radius, $R_\mathrm{P}/R_\star$, in several bins to produce a measured transmission spectrum. 

\begin{figure}[h]
\epsscale{1.2}
\plotone{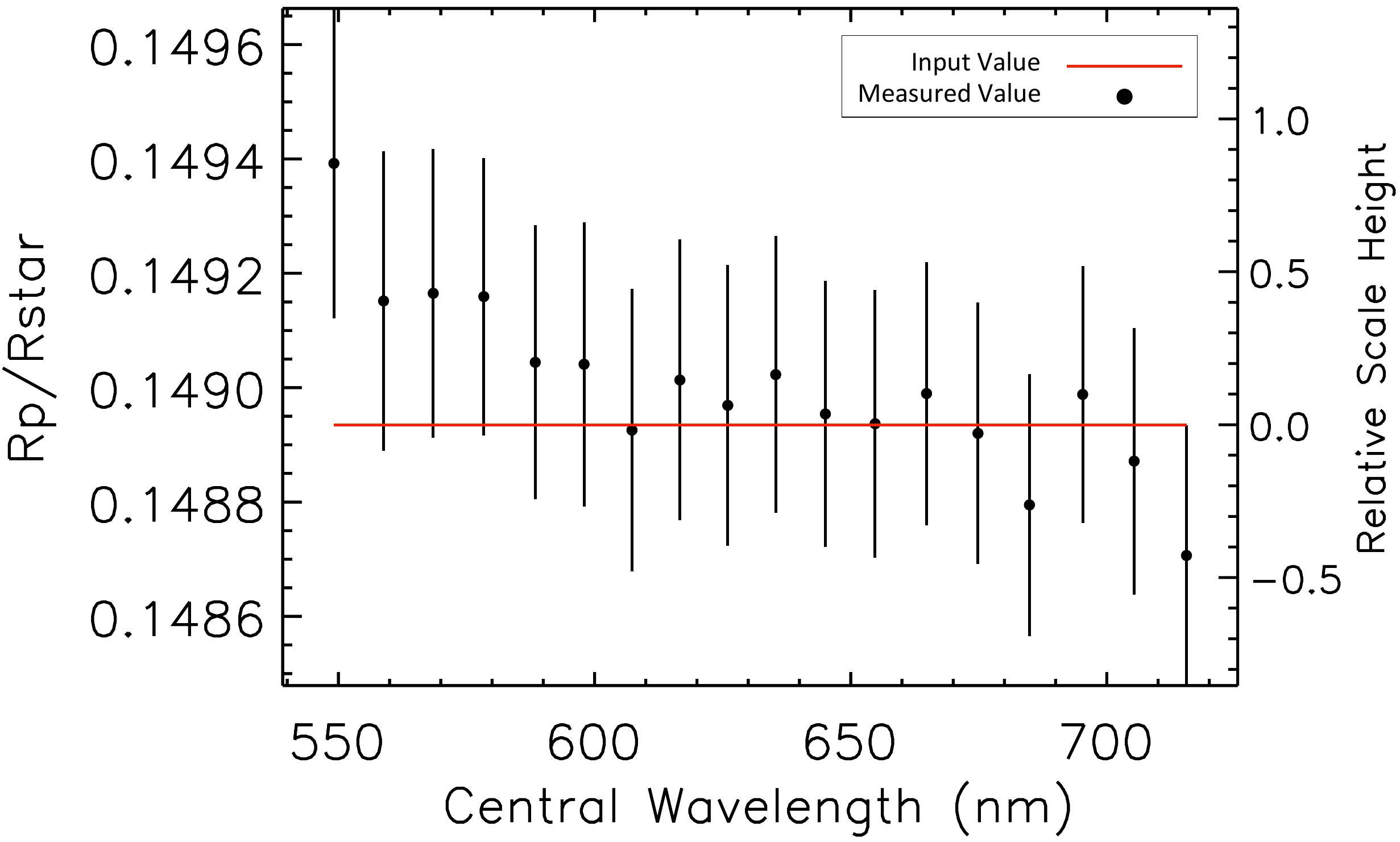}  
\caption{Retrieved values of $R_\mathrm{P}/R_\star$ (black points) compared to an input value (red line) as a function of wavelength. The measured values deviate from the input $R_\mathrm{P}/R_\star$ value more significantly as the bin is moved towards bluer wavelengths, which are closer to the spectral edge. The range of pixels covered here is from 100-600 in the dispersion direction.}
\label{fig_rprstest}
\end{figure}

Figure~\ref{fig_rprstest} shows the results for R150 observation~1, in which any deviation from a constant transit radius is due only to the effect of the stretch systematic. The un-corrected spectral stretch introduces a spectral slope in the transmission spectrum. Given the amplitude of the spurious signal (about 1 atmospheric scale height), this could be mistaken for scattering in the planetary atmosphere. 
The slope is caused by the fact that the flux variation caused by a shift depends on the instrument response function. Where the stellar continuum is relatively flat, such as the 600-650~nm region, the flux changes introduced by a given shift are negligible. As the response function steepens towards bluer wavelengths, however, the introduced flux changes become larger. 

It is therefore critically important to measure and correct for differential spectral shifts where the continuum is a steep function of wavelength. It is also likely to be more important when the target and reference star are not of the same spectral type as they are for WASP-4. In this case, a given shift at a given wavelength could produce significantly different flux variations between the two stars.

As a final test, we repeated the test described above for a variety of bin sizes and found that the deviations between input and measured $R_\mathrm{P}/R_\star$ compared to photon noise were smallest for 10-20~nm bin widths. This is the same range of bin sizes that were found to optimize the rms of the out-of-transit lightcurves in Section~\ref{sec_analysis}. 
This suggests that spectral shifts contribute noticeably to red noise in the transit lightcurves. Importantly, the optimum spectral bin is not the largest possible. This is partly because, while the shift is a smaller fraction of the bin size for larger bins, photon noise is also lower and so the effect of a shift compared to photon noise is higher.

\begin{figure}[h]
\epsscale{1.1}
\plotone{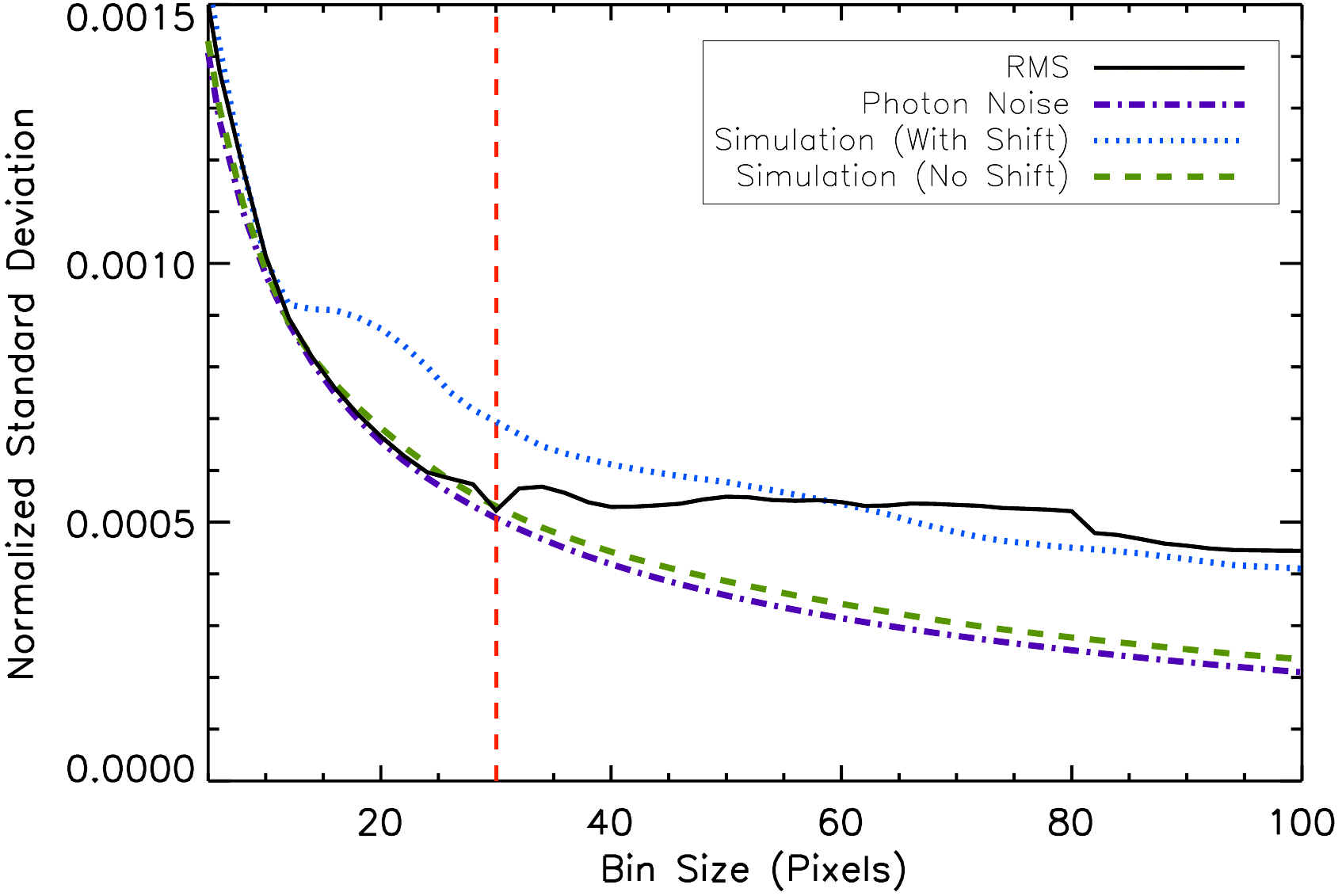}  
\caption{Standard deviation as a function of bin size for a bin centered on the Na~\textsc{i} core in various cases. Shown are the prediction from photon noise (purple dot-dashed), the standard deviation measured in the real data (black solid line) and the standard deviations predicted by the simulation due to only the shift (blue dotted) and in the case where the shift has been corrected (green dashed). All standard deviations are normalized to the out-of-transit flux. Shown with a dashed red line is the bin size used for the majority of the transmission spectrum, which is 30 pixels or 10~nm.}
\label{fig_simtest}
\end{figure}

Figure~\ref{fig_simtest} shows the real measured standard deviation as a function of bin size for the bin around the Na~\textsc{i} feature along with the standard deviation computed using the simulation both with and without injected shifts. While the standard deviation does decrease for larger bins in the simulation, the standard deviation compared to photon noise is smallest at 30 pixels, which corresponds to 10~nm. It is worth noting that the real data are corrected for the spectral shifts and therefore should not be expected to follow the simulation. However, both predict the same optimal bin size, assuming that we restrict ourselves to bins larger than 5~pixels, for which photon noise is prohibitively large. This indicates that the bin size cannot be chosen randomly; thus, optimizing the bin size is crucial step for such a study.
 
\subsection{Noise Sources for the Transmission Spectrum of WASP-4b}
\label{sec_noise}

Before interpreting the transmission spectrum of WASP-4b that was presented in Section~\ref{sec_trans_spec}, we now consider remaining sources of uncertainty not previously discussed.
For each source of uncertainty, we investigated whether it has a significant effect on the transmission spectrum and, if needed, what steps we took to mitigate or quantify its effect. This section includes noise sources specific to GMOS and sources more generally applicable to exoplanet atmosphere observations.

Firstly, we tested the robustness of our cross-correlation procedure detailed in Section~\ref{sec_1d_corrections}. Since each measured cross-correlation value has an associated uncertainty, we wanted to ensure that such uncertainties were not significant when producing the spectral lightcurves and hence the transmission spectrum. To perform the test, we re-reduced all the data but rather than applying the real cross-correlation coefficients, we instead randomly applied 3~$\sigma$ offsets to the cross-correlation coefficients. We then constructed lightcurves, fitted the transits and produced a transmission spectrum using the same methodology used to produce the real transmission spectrum.

We found that the majority of measured transit depths were within 1~$\sigma$ of the values measured using the real cross-correlation coefficients. We can therefore conclude that uncertainty in our cross-correlation has a negligible effect on the transmission spectrum in the majority of bins. For the points where the effect was not negligible, investigation revealed that all of these wavelengths were within strong telluric or stellar absorption bands. For the R150 observations, these compromised bins were the bin located at the stellar H$\alpha$ feature at 656.3~nm and several bins centered on 770~nm, which are within a deep telluric O$_2$ band. We omitted all affected bands from our transmission spectra in Figure~\ref{finalbroadspec}. No transit depths were affected at $> 1 \sigma$ for the B600 observation. 

For the bin around the H$\alpha$ feature, we investigated why the bin was compromised, since variation in that line could be due to transit-correlated H$\alpha$ absorption. We find a larger transit radius in that bin compared to the surrounding bins ($\Delta R_P/R_\star = 0.0015 \pm 0.0004$ for a weighted mean across all observations) although the radii measured for each individual transit are different at the 3-$\sigma$ level. This could potentially be consistent with variable absorption relating to the transit. However, we also note that there was a large group of bad columns in the reddest wing of the H$\alpha$ feature that are omitted, and so our bin does not cover the feature entirely. Combined with the fact that the measured radii in this bin are sensitive to our cross-correlation solution, we choose not to include the result from this bin in our analysis. WASP-4b could, however, benefit from higher-resolution observations around the H$\alpha$ feature to further investigate the possibility of absorption due to the planet transit.
 

We also performed tests on the robustness of our background subtraction. We first tested the degree to which the background could contaminate our resultant transmission spectra by performing the extraction with the background coefficients multiplied by 10$\times$. We then constructed lightcurves and transmission spectra as before and compared the resulting transmission spectrum with the transmission spectrum produced using the real background subtraction coefficients. We found that all $R_P/R_\star$ values in the final transmission spectra deviated by much less than 1-$\sigma$ between the two cases, indicating that the final results are robust to uncertainties in background subtraction and that the contribution of the background to our spectra is negligible.

Additionally, we identified a series of exposures in R150 transit~1 during which there were likely thin clouds, although clouds are not recorded in the night log. In exposures 83-95, the stellar fluxes dropped by a factor of 2 while the background flux remained within the standard deviation measured from the rest of the exposures. The resultant transmission spectrum is unaffected by the inclusion of these exposures at the 0.2~$\sigma$ level and so we included them in the analysis.

Our next set of tests focused on our lightcurve fitting procedure. We first tested the results of using our chosen systematics models in the transit lightcurve fits against the results of using all other parameterizations investigated in Section~\ref{sec_parameterization}.

We found that absolute transit depths were affected at the 4-$\sigma$ level when including some of the observation-long trends, such as functions of CRPA, with the direction of the difference depending on the direction of variation of the associated parameter with time. However, the $\Delta R_P/R_\star$ values that we quote  differ at only the $\sim 0.5~\sigma$ level across all parameterizations. This further reinforces our conclusion that quoting differential transit depths as a function of wavelength rather than absolute transit depths is the best way to proceed.

We also extracted the broad-band transmission spectrum with fits using a linear limb-darkening law with a free coefficient and compared this with our chosen method of using a non-linear law with fixed coefficients. We found that the transmission spectra obtained using both methods were consistent with one another but that uncertainties on measured $\Delta R_P/R_\star$ were 1.2-1.5 times higher when using the free linear coefficient. When examining the residuals of the lightcurve fits, we found that residuals were slightly noisier when using the linear law, suggesting that it does not capture the shape of the transit as well as the non-linear law at our precision.

We then tested whether the 1-nm level uncertainty in our wavelength solution could affect our fixed limb-darkening coefficients in the transit lightcurve fits enough to affect measured $\Delta R_P/R_\star$ measurements. To test this, we computed, for each spectral bin, limb-darkening coefficients for our actual wavelengths and for wavelengths at $\pm 1 \sigma$ using the same stellar atmosphere model as used in Section~\ref{sec_binned_LC}.

We produced two model transit lightcurves using the system parameters of WASP-4 for each of our spectral bins. Each lightcurve used a different set of limb-darkening coefficients based on the two different possible wavelength values for that bin. For each lightcurve, we fitted the transit and measured the difference in retrieved $R_\mathrm{P}/R_\star$ between the two sets. The change in measured $R_\mathrm{P}/R_\star$ was less than $1 \sigma$ for all our spectral bins and so we conclude that our $R_\mathrm{P}/R_\star$ measurements are reliable with respect to small uncertainties in the wavelength solution.

Finally, we modeled the possible wavelength-dependent effect of stellar spots on the transmission spectrum. This is because WASP-4 is a moderately active star with observed 1~\% variations in transit depth in red optical and $z$ bands \citep{2013MNRAS.434...46H}. To model the potential contribution of stellar spots to the transmission spectrum, we followed the procedures detailed in \citet{2011A&A...526A..12D}, \citet{sing11} and \citet{2014ApJ...791...55M} to model the non-spotted stellar surface and stellar spots as blackbodies of different temperatures. Assuming a conservative 2~\% variability in the optical, the contribution from stellar activity is within our uncertainties in the transmission spectrum regardless of spot temperature. This is consistent with our observation that the transmission spectra agree well between R150 observations~1,~2 and~3, which were unlikely to be observed at the same activity level. 

\section{Discussion and Interpretation}
\label{sec_discussion}

\subsection{Transit Parameters}
\label{sec_syspar}

In Table~\ref{system_params_table}, we show the results of fitting each transit separately for planetary system parameters. This enables us to check whether the different observations produce consistent results. We find that the values of $i$ and $a/R_\star$ are not consistent across the multiple observations at the 2-3 sigma level. 
We find, however, that the fitted values of $R_\mathrm{P}/R_\star$ agree within 1-$\sigma$ for each of the R150 observations. In addition, we have shown in Section~\ref{sec_analysis} that varying the system parameters when fitting for the wavelength-dependent transit radii affects only the absolute transit depths and not the relative transit depths. We are therefore confident about combining these results to produce the transmission spectrum.



We additionally performed a fit using all 4 transits simultaneously. In this method, we fitted all 4 transits with a single value of $a/R_\star$ and $i$. For all R150 transits, we used wavelengths bluer than 700~nm only, so that we could fit a single value of $u$ and $R_P/R_\star$ for all 3 R150 transits. We fitted the B600 transit with a separate value of $u$ and $R_P/R_\star$. Each transit was also fitted with a separate central transit time and separate parameters governing the systematics and baseline stellar flux, using the models described in Section~\ref{sec_whitelight}. 
We assumed a fixed period of 1.33823204~days \citep{2013MNRAS.434...46H}.
The retrieved parameters from the joint fit are given in Table~\ref{jointfit_params}. The individual ephemerides for each transit are identical to those given in Table~\ref{system_params_table}. We find that the timing variations between the 4 transits are consistent with the propagated uncertainty on the orbital period determined by \citet{2013MNRAS.434...46H}.

\begin{table}[h]
\centering
\begin{tabular}{lc}
\hline
\hline
Parameter & Value \\
\hline
$i$ (deg) & $87.63^{+0.32}_{- 0.28}$ \\
$a/R_\star$ & $5.41^{+ 0.031}_{- 0.030}$ \\
$R_P/R_\star$ (550-700~nm) & $0.1566^{+ 0.0004}_{-0.0004}$ \\
$R_P/R_\star$ (400-650~nm) & $0.1537^{+ 0.0008}_{-0.0008}$ \\
$u$ (550-700~nm) & $0.53^{+0.008}_{-0.009}$ \\
$u$ (400-650~nm) & $0.68^{+0.016}_{-0.017}$ \\
\hline
\end{tabular}
\caption{Fitted parameters for the WASP-4 system, using all 4 transits to jointly fit single values of $i$ and $a/R_\star$ to all transits and a value of $R_P/R_\star$ and $u$ for each bandpass.}
\label{jointfit_params}
\end{table}


We are unable to improve the uncertainties on the fitted system parameters over those found by \citet{2013MNRAS.434...46H}, since that study employed a combined analysis of 38 transit lightcurves. In order, therefore, to compare our derived $R_P/R_\star$ values with the literature, we repeated the joint fit, but this time fixed the $i$ and $a/R_\star$ values to those found by \citet{2013MNRAS.434...46H} for their analysis of 38 transits. We obtained $R_P/R_\star$ values of $0.1549 \pm 0.0002$ for the 550-700~nm range and $0.1523 \pm 0.0006$ for the 400-650~nm range. The derived value of $R_P/R_\star$ by \citet{2013MNRAS.434...46H} was $0.15445 \pm$ 0.00025, and the 38 transits were primarily composed of red to near-IR bands, with some visible and green transits also included. Our fitted value in the 550-700~nm range is consistent with that determined by \citet{2013MNRAS.434...46H}, which is expected since we cover a similar bandpass as the majority of their transit observations. The fitted value for the blue wavelengths, however, is not consistent at the 3-$\sigma$ level with the previous result.  

When fitting for $i$ and $a/R_\star$, our derived values of $R_P/R_\star$ differ again from those found when fixing the other system parameters to those in \citet{2013MNRAS.434...46H} (Table~\ref{jointfit_params}). In addition, our fitted value for $i$ differs at the 2-$\sigma$ level from that found by \citet{2013MNRAS.434...46H} although our fitted value for $a/R_\star$ agrees within the uncertainties. 
In practice, stellar activity, and more specifically stellar spots, can affect the transit parameter retrieval. Indeed,
 WASP-4 shows moderate flux variations due to activity, and spot crossings have been identified in 15-30~\% of previous $z$-, $I$- and $R$-band observations \citep{2013MNRAS.434...46H}. 
However, the agreement between the independently-measured $R_P/R_\star$ for each R150 observation suggests that stellar activity does not significantly affect the measured transit radii, and we have found in Section~\ref{sec_analysis} that our relative transit depths are not sensitive to the absolute system parameters.

\subsection{Low-Resolution Transmission Spectrum and Atmospheric Models}
\label{sec_disc_lowres}

To broadly characterize the atmospheric transmission spectrum of WASP-4b, we constructed three different models and compared them with our observed transmission spectrum. The models broadly describe 1) an aerosol-free forward model from \citet{fortney10}, 2) an atmospheric transmission spectrum dominated by small scattering grains and 3) an atmospheric transmission spectrum dominated by large-scattering grains.

The aerosol-free forward model is for the WASP-4 system and includes a self-consistent treatment of radiative transfer and chemical equilibrium of neutral and ionized species \citep{fortney10}. Chemical mixing ratios and opacities assume solar metallicity and local chemical equilibrium accounting for condensation and thermal ionization but no photoionization \citep{lodders99,loddersfegley02,lodders02,visscher06, lodders09,freedman08}. For the small-grain dominated model, we used a Rayleigh scattering opacity. Finally, for the large-grain scattering model, we simply used a flat line for the transmission spectrum. 

To compare the models with the data, we first flux-weighted and binned them to the resolution of the data. We then fitted each model to the transmission spectrum allowing the model to shift up and down in absolute $R_\mathrm{P}/R_\star$ while also allowing a relative shift, $\Delta z$, between the B600 and R150 transmission spectra. The $\Delta z$ parameter is necessary because we only measure relative transit depths in each grating and so the two sets of data can move up or down with respect to one another. While we do measure absolute transit depths from the white light curves, these are sensitive to the model chosen for the systematics, as mentioned in Section~\ref{sec_noise}, and so we use only the relative transit depths for each grating for the fit along with the $\Delta z$ parameter. The result is a fit with 2 free parameters for each of the atmospheric models tested. 

\begin{figure*}
\centering
\epsscale{1.0}
\plotone{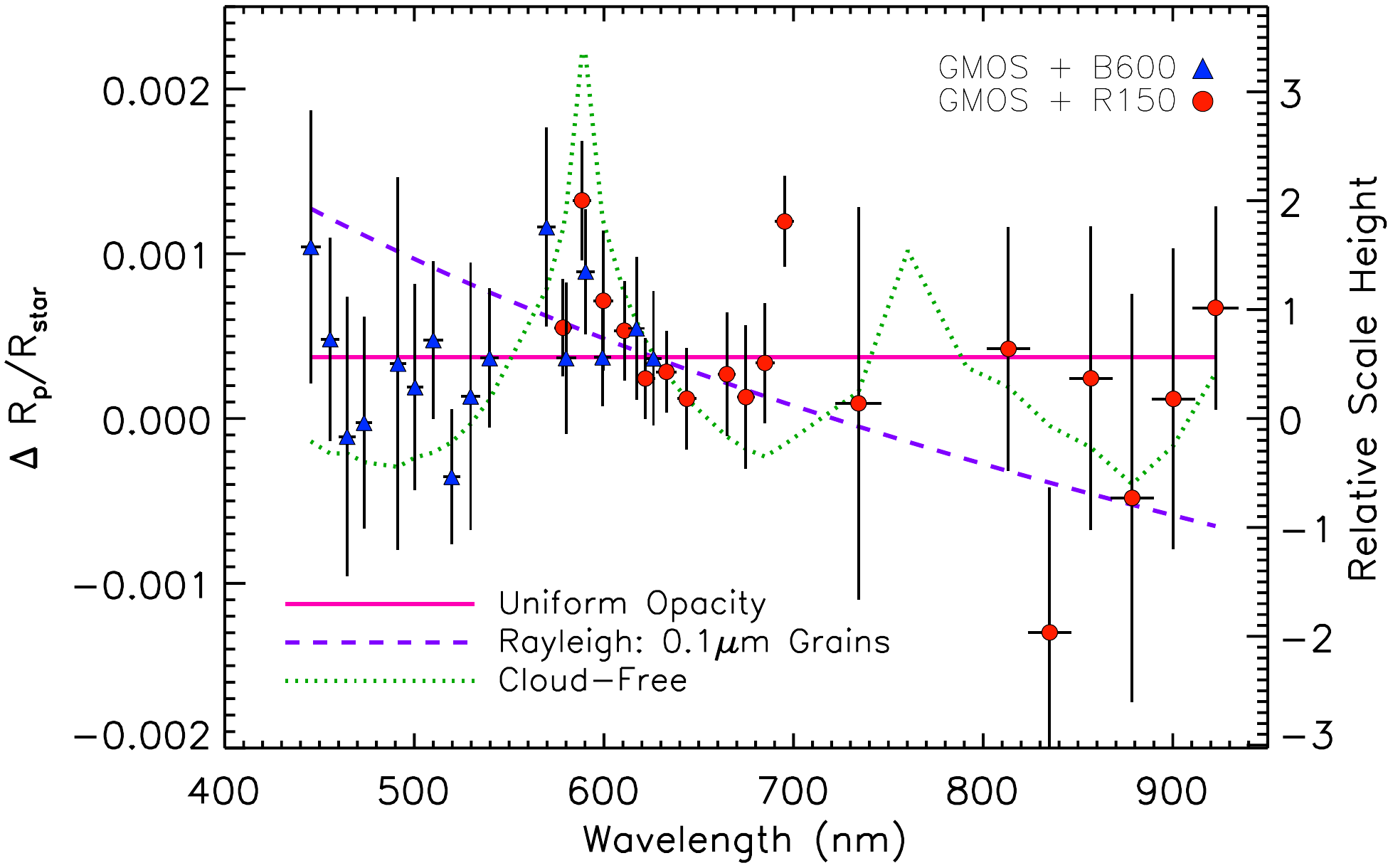}   
\caption{Transmission spectrum of WASP-4b from GMOS B600 data (blue triangles) and GMOS R150 data (red circles). The absolute levels of the B600 radii with respect to the R150 radii are plotted using the best-fitting $\Delta z$ parameter from the large-grain-dominated atmosphere fit. The uncertainties on the data points are the uncertainties on the wavelength-dependent $R_P/R_\star$ only, and do not include the uncertainty on the absolute transit depth or the $\Delta z$ parameter.
Also shown are the cloud-free forward atmospheric model from \citet{fortney10}, an atmosphere dominated by scattering from small-particle grains and an atmosphere dominated by scattering from large-particle grains. 
Models are binned to the resolution of the data. Note that the lower precision redward of 700~nm is due to residual fringing and also because we were able to use only transit~3 for this wavelength range. Blueward of 700~nm, we combined the results of transits~1,~2 and~3.}
\label{finalspec_and_models}
\end{figure*}

We firstly fitted each model to the whole dataset. Then, due to the lower precision at longer wavelengths, we also tried fitting each model to only the data blueward of 700~nm. The goodnesses of fit of each model in both cases are shown in Table~\ref{model_chisq}. The model fits to all data points are shown in Figure~\ref{finalspec_and_models}. Note that there are no data points around the predicted K~\textsc{i} feature due to telluric contamination (see Section~\ref{sec_noise}). 

\begin{table}[h]
\centering
\begin{tabular}{ccc}
\hline
\hline
\multicolumn{3}{c}{Fit to All Data Points} \\
Model & Reduced $\chi^2$ & $\Delta z$ \\
\hline
Large-Grain Dominated & 1.25 &  $-0.00019 \pm 0.00009$ \\  
Small-Grain Dominated & 1.59 & $0.00025 \pm 0.00009$ \\ 
Cloud-Free Atmosphere & 2.53 & $0.00012 \pm 0.00009$ \\
\hline
\end{tabular}
\begin{tabular}{ccc}
\multicolumn{3}{c}{Fit to Only Data Points Blueward of 700~nm} \\
Model & Reduced $\chi^2$ & $\Delta z$ \\
\hline
Large-Grain Dominated & 1.00 &  $-0.00017 \pm 0.0001$ \\ 
Small-Grain Dominated & 1.40 & $0.00019 \pm 0.0001$ \\ 
Cloud-Free Atmosphere & 2.76 & $0.00011 \pm 0.0001$ \\
\hline
\end{tabular}
\caption{Reduced $\chi^2$ values for atmospheric model fits to the transmission spectrum of WASP-4b. The top table shows fits to all data points (34 degrees of freedom) and the bottom table shows fits to only the data points blueward of 700~nm (25 degrees of freedom). Higher values of $\Delta z$ indicate that the blue radii are higher with respect to the red radii. A $\Delta z$ of zero would indicate that the average radii for the B600 and R150 radii are equal.}
\label{model_chisq}
\end{table}

The model fits indicate that the low-resolution optical transmission spectrum of WASP-4b is best fit with a uniform opacity or flat line. Such a spectrum is consistent with large grain scattering, which could be due to aerosols.
Based on the equilibrium temperature of WASP-4b, 1700~K, several species could form condensates including MnS, MgSiO$_3$, Fe and Al$_2$O$_3$ \citep{morley12}. 
Future observations redward of 10~$\mu$m with JWST would provide the best constraint on the scattering species through its absorption features \citep{2017MNRAS.464.4247W}. 

We also searched for the possible presence of TiO in the transmission spectrum of WASP-4b, which has long been a possible atmospheric constituent of hot Jupiters \citep{fortney08,desert08,2015A&A...575A..20H}. We again used the models described in \citet{fortney10} to produce a forward model for the WASP-4 system, this time containing TiO opacities. We then fit the model to our observed transmission spectrum using the same procedure as for the other models investigated. 
The result was that an aerosol-free atmosphere containing a solar abundance of TiO gives a $\chi^2_\nu$ of 1.94 for 34 DOF when fit to all the data points, compared to $\chi^2_\nu$ of 1.25 for the large-grain dominated model. We therefore conclude that there is no significant evidence of TiO in the upper atmosphere of WASP-4b. 

Given the inhomogeneous day-night thermal redistribution observed by \citet{2011A&A...530A...5C} in WASP-4b, it is possible that a cold-trap for TiO exists on the planet's night side as described by \citet{spiegel09} and \citet{parmentier12}. So far, the only exoplanet atmospheres in which TiO has been detected are WASP-121b and WASP-33b, which are 500-700~K hotter than WASP-4b \citep{2015ApJ...806..146H,2016ApJ...822L...4E}. This could suggest that TiO only exists in the atmospheres of the hottest planets. 

We additionally note that the $\Delta z$ values for all of the atmospheric model fits are much smaller than the difference between the measured $R_P/R_\star$ in the blue and red gratings given in Table~\ref{jointfit_params}, even when taking into account the associated uncertainties. In Figure~\ref{finalspec_and_models}, the radii measured with the B600 and R150 gratings agree well where the two gratings overlap, indicating that the fitted value of $\Delta z$ is likely realistic. This further suggests that the absolute $R_P/R_\star$ measurements should be taken with caution, and that only the $\Delta R_P/R_\star$ measurements in each grating are robust. 

\subsection{The Search for High-Resolution Na~\textsc{i} Signal}

At low resolution, we do not have high enough precision to conclusively detect or rule out the presence of Na~\textsc{i} in the transmission spectrum of WASP-4b. We now investigate whether we can detect evidence of Na~\textsc{i} in the atmosphere of WASP-4b using the higher-resolution bins discussed in Section~\ref{Na_analysis}. 

\begin{figure}[h]
\epsscale{1}
\plotone{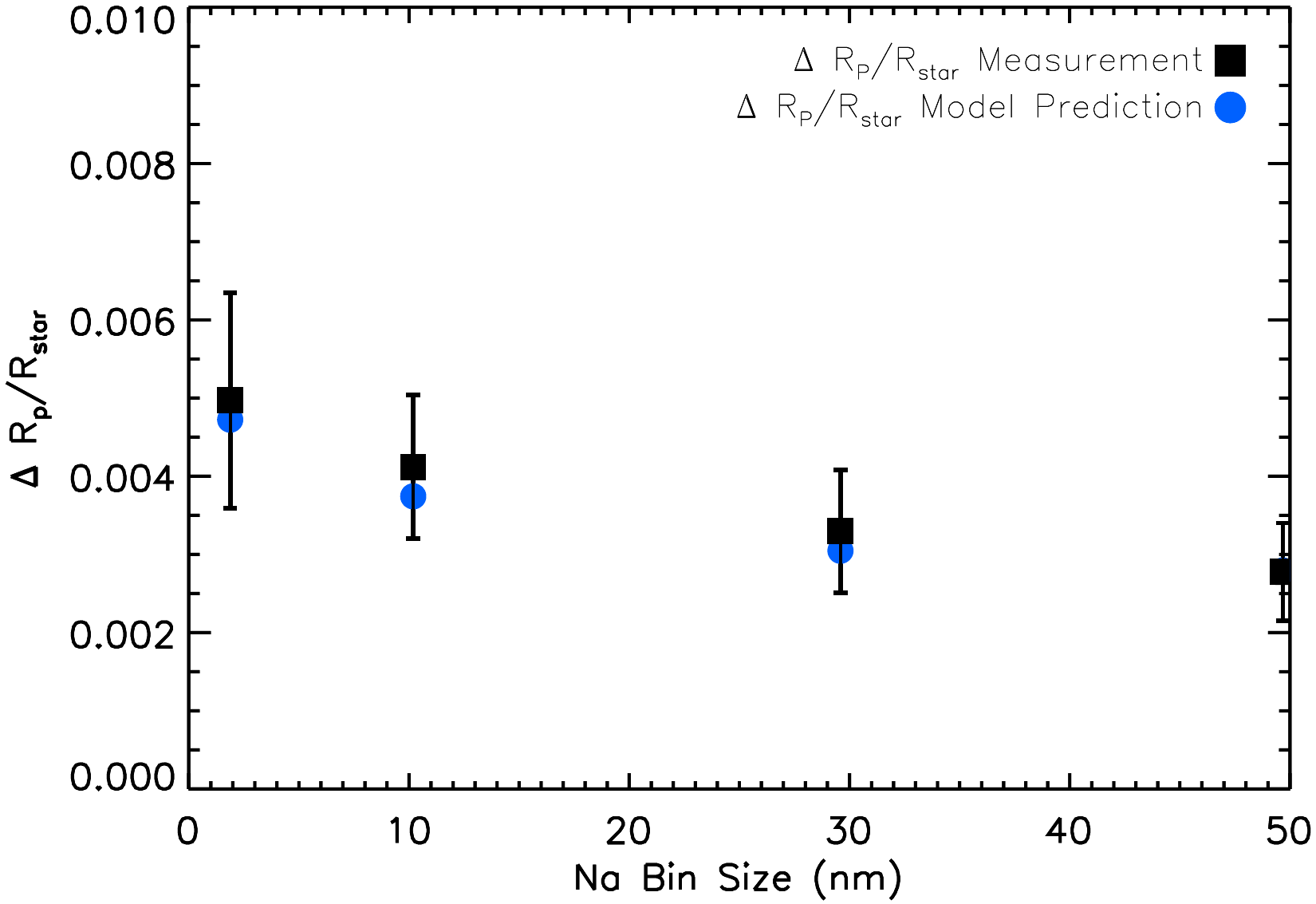}
\caption{Differential $R_\mathrm{P}/R_\star$ measurements between a bin centered on the Na~\textsc{i} doublet and a continuum bin from 675-685~nm (black squares). This is a weighted mean of all observations. Overplotted with blue circles are the values obtained from an atmospheric model assuming an aerosol-free model with no TiO opacity \citep{fortney10}.}
\label{IADNaMean}
\end{figure}

Figure~\ref{IADNaMean} shows the measured $\Delta R_\mathrm{P}/R_\star$ between a bin centered on the Na~\textsc{i} feature core and the continuum bin at 675-685~nm for a variety of bin sizes around Na~\textsc{i}. The values shown are the weighted mean across all observations, which are all consistent with one another. Model values are also shown assuming the aerosol free model with no TiO opacity, in which the Na~\textsc{i} feature should be prominent in the transmission spectrum (e.g. Figure~\ref{finalspec_and_models}).

There is a decrease in $\Delta R_\mathrm{P}/R_\star$ for larger bin sizes around Na~\textsc{i}, which is consistent with the aerosol-free model. As discussed in \citet{huitson12}, this measurement is not sensitive to the profile of the Na~\textsc{i} feature (narrow core only vs. broad with wings) but the decrease in $\Delta R_\mathrm{P}/R_\star$ as a function of increasing bin size could be consistent with the presence of Na in the atmosphere of WASP-4b. 

\section{Conclusions}

We have observed the complete optical transmission spectrum of the hot Jupiter WASP-4b from the ground using Gemini GMOS South. This is the first result from a survey of 9 hot Jupiter atmospheres using the same instrument and technique. We have also presented the data reduction pipeline, and made a comprehensive study of artifacts of the  GMOS instruments for transmission spectroscopy of exoplanet atmospheres.

We find that the transmission spectrum of WASP-4b is dominated by a uniform opacity, consistent with high altitude aerosols, or large $\sim 1$~$\mu$m grain sizes rather than small grains of $\sim~0.1$~$\mu$m. This finding is consistent with many published hot Jupiter transmission spectra and is not surprising since aerosol cover has been seen in planets with hotter equilibrium temperatures than WASP-4b, where it is expected that fewer species could condense. 

We have searched for the presence of Na~\textsc{i} in the atmosphere of WASP-4b by looking at the highest resolution provided by our data (in 2~nm bins) and found a signal consistent with the presence of Na~\textsc{i} at the 2-$\sigma$ level, but cannot definitely conclude on the detection of this element. 
We were unable to measure the transmission spectrum around the predicted K~\textsc{i} feature due to telluric contamination.

WASP-4 could potentially benefit from higher-resolution observations around the Na~\textsc{i} doublet in order to detect the presence or absence of this species. 
In the longer term, observations with JWST redward of 10~$\mu$m would provide the best constraint on the aerosol species through their absorption features. 

We find no evidence of TiO absorption in the optical transmission spectrum, which is not surprising since the only hot Jupiter in which TiO has been found is 700~K hotter than WASP-4b. A large day-night contrast indicates that TiO may be cold-trapped out of the atmosphere of WASP-4b.
While we see evidence for a variation in central transit time in our multiple-epoch transit lightcurves, these are below the level of uncertainty on the orbital period.

For two out of our three R150 observations, significant and time-varying fringe patterns prevented us from obtaining useful transmission spectral data at wavelengths redder than 700~nm. We were, however, able to confirm excellent agreement between the wavelength-dependent transit depths observed in all three observations in the range 570-700~nm. Since these three observations span a time period of two years, this indicates a high reliability in the Gemini GMOS setup. We additionally obtained one transit after the GMOS detector was upgraded, and can confirm that the fringe amplitude has been reduced by 10~\%, enabling transmission spectroscopy redward of 700~nm.

We conducted a thorough investigation of the dominant noise sources in the Gemini GMOS transmission spectra for our whole survey dataset. For broad-band transmission spectroscopy, we find that the dominant instrumental systematics are due primarily to optical variations, which cause a time- and wavelength-dependent displacement of stellar spectra on the detector during an observation. The effect is partly accounted for by a model of differential atmospheric refraction but there is an additional component which cannot be accounted for. We developed and tested an empirical model for removing the additional effect. We also see a discretization of points at the 10-count level which is likely electronic but which does not dominate our uncertainties. 

We have found that Gemini GMOS is capable of optical transmission spectroscopy at the precision of space-based instruments, once the spectral stretching is accounted for. We intend to make our pipeline public so that the community has access to the tools that we used to account for the systematic noise sources seen in the GMOS data presented here.

\acknowledgments

Based on observations obtained at the Gemini Observatory (acquired through the Gemini Observatory Archive and Gemini Science Archive), which is operated by the Association of Universities for Research in Astronomy, Inc. (AURA), under a cooperative agreement with the NSF on behalf of the Gemini partnership: the National Science Foundation (United States), the National Research Council (Canada), CONICYT (Chile), Ministerio de Ciencia, Tecnolog\'{i}a e Innovaci\'{o}n Productiva (Argentina), and Minist\'{e}rio da Ci\^{e}ncia, Tecnologia e Inova\c{c}\~{a}o (Brazil).

Based on Gemini observations obtained from the National Optical Astronomy Observatory (NOAO) Prop. IDs: 2012B-0398 and 2011B-0387; PI: J-.M D\'{e}sert. 

The research leading to these results has received funding from the European Research Council (ERC) under the European Union's Horizon 2020 research and innovation programme (grant agreement no. 679633; Exo-Atmos). This material is based upon work supported by the National Science Foundation (NSF) under Grant No. AST-1413663, and supported by the NWO TOP Grant Module 2 (Project Number 614.001.601).

This research has made use of NASA's Astrophysics Data System.

\textsc{iraf} is distributed by the NOAO, which is operated by AURA under a cooperative agreement with the NSF.

\software{IRAF \citep{1986SPIE..627..733T, 1993ASPC...52..173T}, diff\_atm\_refr.pro (http://www.eso.org/gen-fac/pubs/astclim/lasilla/diffrefr.html), emcee \citep{2013PASP..125..306F}, ATLAS (Kurucz 1993, available at http://kurucz.harvard.edu/stars/hd189733), \\ mpfit.pro \citep{markwardt09}, Time Utilities \citep{2010PASP..122..935E}, BATMAN \citep{2015PASP..127.1161K}.}

\bibliographystyle{apj}
\bibliography{wasp4gmos}

\appendix

\section{Transit Lightcurves}
\label{sec_trans_lc}

\begin{figure*}[h]
\centering
\epsscale{1}
\includegraphics[width=17cm]{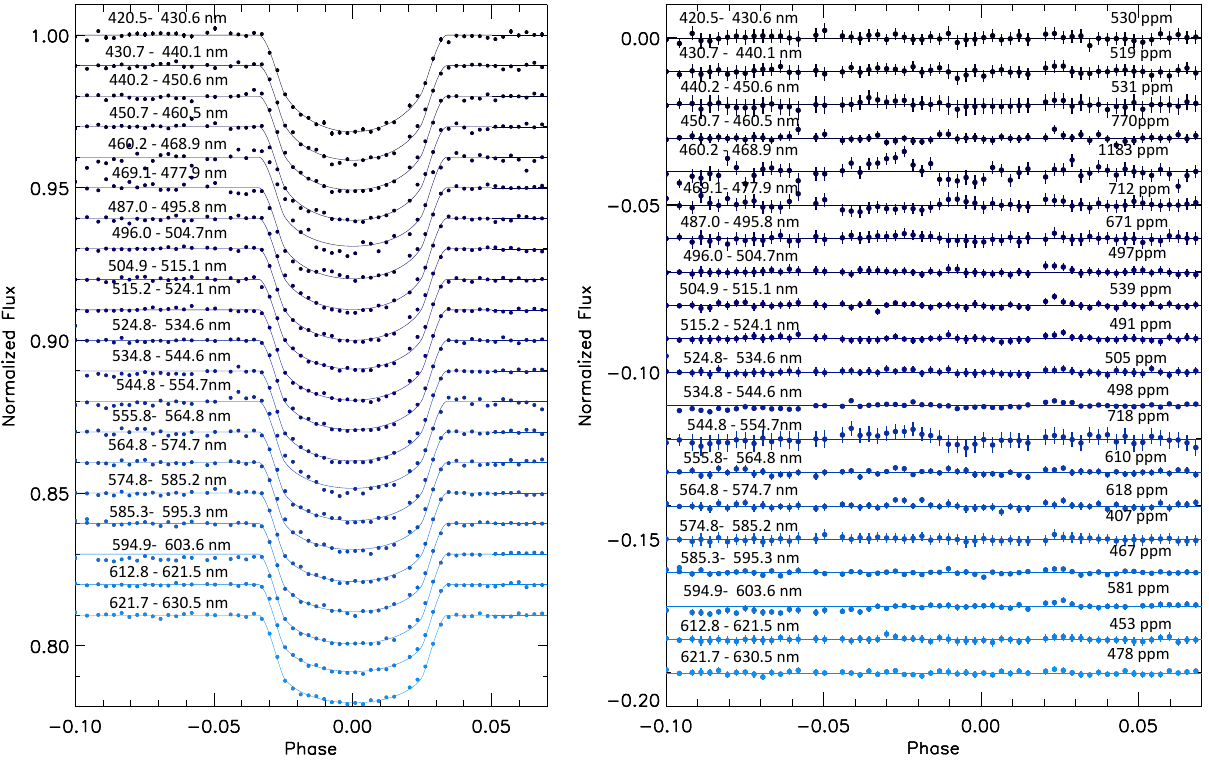}  
\caption{\textit{Left: }GMOS B600 spectral light curves in each spectral bin after removal of systematics and normalization, overplotted with the best-fitting transit models from \citet{mandelagol02}. The spectral lightcurves are plotted with longer-wavelength bins at the bottom, and each lightcurve has an arbitrary flux offset for clarity. \textit{Right: }Corresponding residuals for each spectral lightcurve fit, with uncertainties rescaled with the $f$ parameter. Outliers have been clipped but the points missing around phase -0.045 and phase 0.02 are not present in the original observation. Some bands display obvious residual trends. These are close to bad columns and so are not included in the analysis.}
\label{fig_b600_spectralLC}
\end{figure*}

\begin{figure*}[h]
\centering
\epsscale{1}
\includegraphics[width=17cm]{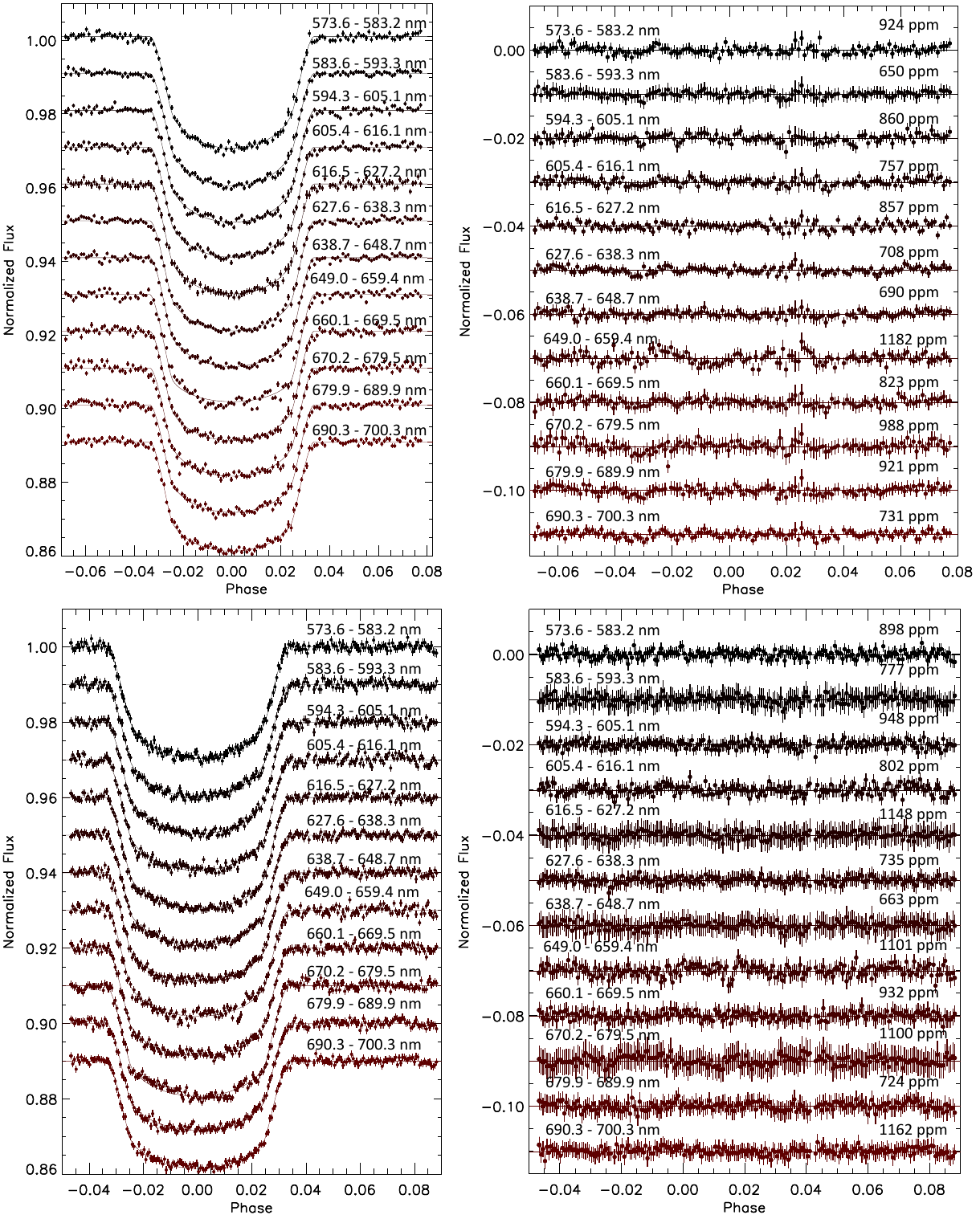}  
\caption{GMOS R150 spectral light curves and residuals for observations~1 and~2. Observation~1 is at the top and observation~2 is at the bottom. Transit lightcurves are shown after removal of systematics and normalization, overplotted with the best-fitting transit models from \citet{mandelagol02}. In each case, the spectral lightcurves are plotted with longer-wavelength bins at the bottom, and each lightcurve has an arbitrary flux offset for clarity. Photometric uncertainties have been rescaled with $f$ in the residuals plots.}
\label{fig_r150_spectralLC1}
\end{figure*}

\begin{figure*}[h]
\centering
\epsscale{1}
\includegraphics[width=18cm]{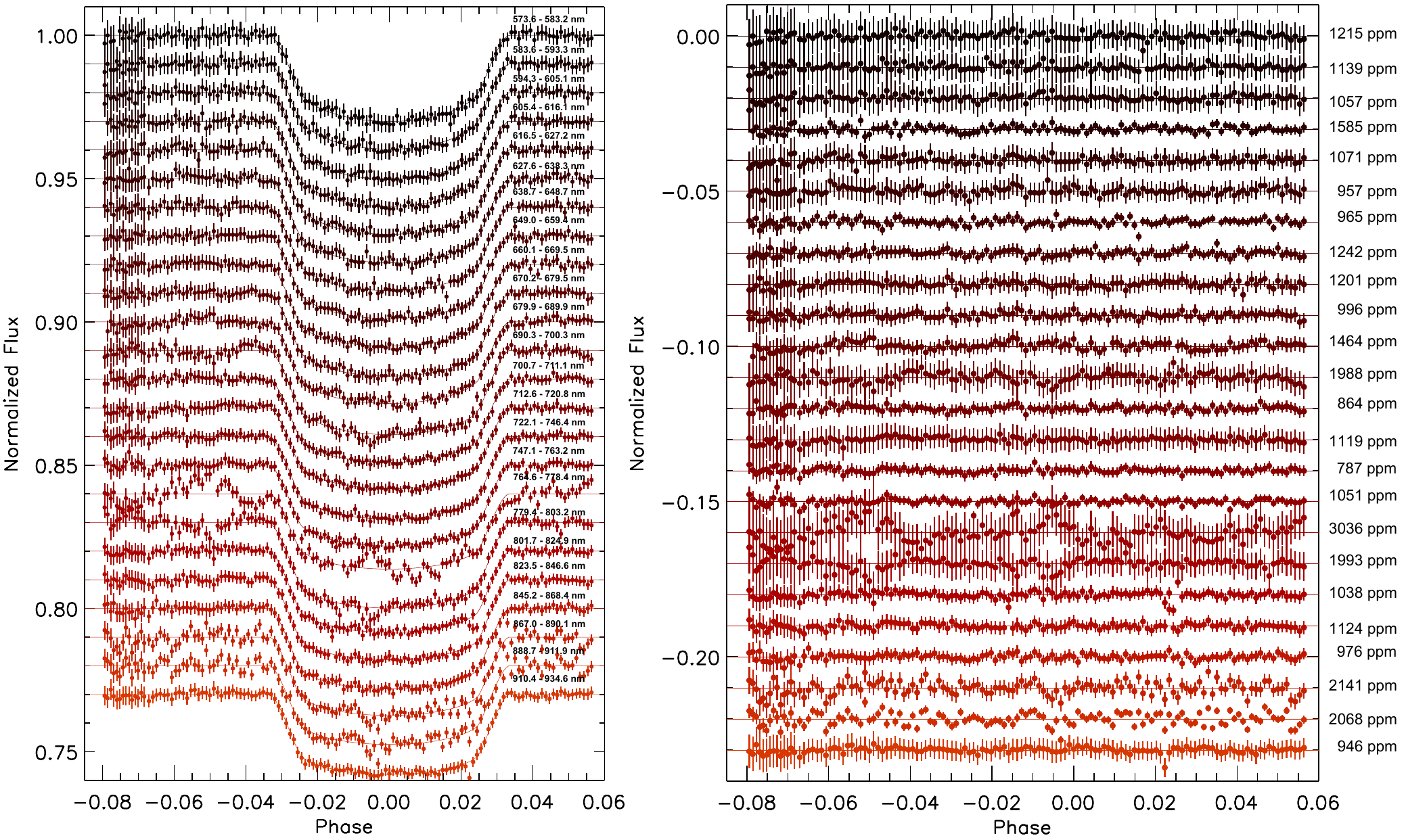}
\caption{GMOS R150 spectral light curves and residuals for observation~3. Transit lightcurves are shown after removal of systematics and normalization, overplotted with the best-fitting transit models from \citet{mandelagol02}. In each case, the spectral lightcurves are plotted with longer-wavelength bins at the bottom, and each lightcurve has an arbitrary flux offset for clarity. Photometric uncertainties have been rescaled with $f$ in the residuals plot.}
\label{fig_r150_spectralLC2}
\end{figure*}

\clearpage
\section{Tabulated Values of $\Delta R_P/R_\star$ for the Transmission Spectrum}

\begin{table}[h]
\centering
\begin{tabular}{ccccc}
\hline
\hline
Wavelength (nm) & $\Delta R_\mathrm{P}/R_\star$ & $c_2$ & $c_2$ & $c_3$ \\
\hline
 440.2   - 450.6    & $ 0.0010 \pm 0.0008   $ & 0.8736  &   -0.0558  &   0.0046 \\
 450.7   - 460.5    & $ 0.0005 \pm 0.0006   $ & 1.0324   &  -0.0836   &   -0.1032 \\
 460.2   - 468.9    & $-0.0001 \pm 0.0008   $ & 1.0746   &   -0.1968  &   -0.0456 \\
 469.1   - 477.9    & $-0.0000 \pm 0.0006   $ & 1.1345   &   -0.3017  &   -0.0120 \\
 487.0   - 495.8    & $ 0.0003 \pm 0.0011   $ & 1.1800   &   -0.4161  &    0.0377 \\
 496.0   - 504.7    & $ 0.0002 \pm 0.0006   $ & 1.1755   &   -0.4430  &    0.0597 \\
 504.9   - 515.1    & $ 0.0005 \pm 0.0005   $ & 1.1324   &   -0.4114  &    0.0564 \\
 515.2   - 524.1    & $-0.0004 \pm 0.0004   $ & 1.2134   &   -0.5583   &    0.1163 \\
 524.8   - 534.6    & $ 0.0001 \pm 0.0008   $ & 1.2221   &   -0.6129   &    0.1473 \\
 534.8   - 544.6    & $ 0.0004 \pm 0.0004   $ & 1.2605   &   -0.6094   &    0.1183 \\
 564.8   - 574.7    & $ 0.0012 \pm 0.0006   $ & 1.2964    &  -0.7170   &    0.1673\\
 574.8   - 585.2    & $ 0.0004 \pm 0.0005   $ & 1.3019   &   -0.7431   &    0.1777\\
 585.3   - 595.3    & $ 0.0009 \pm 0.0004   $ &  1.3346   &  -0.7988    &  0.1994 \\
 594.9   - 603.6    & $ 0.0004 \pm 0.0003   $ & 1.3407    & -0.8506   &   0.2410 \\
 612.8   - 621.5    & $ 0.0005 \pm 0.0004   $ & 1.3457   &   -0.9091   &    0.2632 \\
 621.7   - 630.5    & $ 0.0004 \pm 0.0004   $ & 1.3560    &  -0.8905   &    0.2388 \\
 \hline
  573.6   - 583.2    & $ 0.0006 \pm 0.0003   $ & 1.3162 &    -0.7618   &   0.1865 \\
 583.6   - 593.3    & $ 0.0013 \pm 0.0004   $ & 1.3337   &  -0.7976    &  0.1983\\
 594.3   - 605.1    & $ 0.0007 \pm 0.0004   $ & 1.3377    & -0.8109   &   0.2000 \\
 605.4   - 616.1    & $ 0.0005 \pm 0.0003   $ & 1.3530   &  -0.8870   &   0.2422 \\
 616.5   - 627.2    & $ 0.0002 \pm 0.0002   $ & 1.3487   &  -0.8700   &   0.2298 \\
 627.6   - 638.3    & $ 0.0003 \pm 0.0002   $ & 1.3416   &  -0.8937   &   0.2467 \\
 638.7   - 648.7    & $ 0.0001 \pm 0.0003   $ & 1.3649   &  -0.9435   &   0.2656 \\
 660.1   - 669.5    & $ 0.0003 \pm 0.0004   $ & 1.3702   &  -0.9692   &   0.2777 \\
 670.2   - 679.5    & $ 0.0001 \pm 0.0004   $ & 1.3748   &  -0.9920   &   0.2918 \\
 679.9   - 689.9    & $ 0.0003 \pm 0.0004   $ & 1.3664   &  -0.9943   &   0.2942 \\
 690.3   - 700.3    & $ 0.0012 \pm 0.0003   $ & 1.3687   &   -1.0101    &  0.3026 \\
 722.1   - 746.4    & $ 0.0001 \pm 0.0012   $ & 1.3632   &   -1.0737  &    0.3364 \\
 801.7   - 824.9    & $ 0.0004 \pm 0.0009   $ & 1.3678   &   -1.1198   &   0.3599\\
 823.5   - 846.6    & $-0.0013 \pm 0.0012   $ & 1.3452    &  -1.1189    &  0.3616\\
 845.2   - 868.4    & $ 0.0002 \pm 0.0009   $ & 1.3538   &   -1.1541    &  0.3798\\
 867.0   - 890.1    & $-0.0005 \pm 0.0006   $ & 1.3630   &   -1.2032    &  0.4057 \\
 888.7   - 911.9    & $ 0.0001 \pm 0.0009   $ & 1.3409    &  -1.1863    &  0.4026 \\
 910.4   - 934.6    & $ 0.0007 \pm 0.0006   $ & 1.3703 & -0.9688 & 0.27739 \\
 \hline
\end{tabular}
\caption{Final transmission spectrum of WASP-4b from all observations.}
\label{table_final_transspec}
\end{table}

\begin{table*}
\centering
\begin{tabular}{cccc}
\hline
\hline
Wavelength (nm) & $\Delta R_\mathrm{P}/R_\star$ (obs 1) & $\Delta R_\mathrm{P}/R_\star$ (obs 2) & $\Delta R_\mathrm{P}/R_\star$ (obs 3) \\ 
 \hline
  573.6  -  583.2   & $ 0.0000\pm  0.0005  $ & $ 0.0008 \pm 0.0008  $ & $ 0.0006 \pm 0.0006 $ \\
 583.6  -  593.3   & $ 0.0007\pm  0.0006  $ & $ 0.0015 \pm 0.0005  $ & $ 0.0010 \pm 0.0004 $ \\
 594.3  -  605.1   & $ 0.0002\pm  0.0007  $ & $ 0.0006 \pm 0.0006  $ & $ 0.0006 \pm 0.0004 $ \\
 605.4  -  616.1   & $ 0.0007\pm  0.0005  $ & $ 0.0002 \pm 0.0005  $ & $ 0.0001 \pm 0.0005 $ \\
 616.5  -  627.2   & $ -0.0001\pm  0.0004  $ & $ 0.0002 \pm 0.0008  $ & $ 0.0001 \pm 0.0005 $ \\
 627.6  -  638.3   & $ -0.0002\pm  0.0004  $ & $ 0.0002 \pm 0.0011  $ & $ 0.0004 \pm 0.0005 $ \\
 638.7  -  648.7   & $ 0.0002\pm  0.0005  $ & $ -0.0005 \pm 0.0008  $ & $ 0.0002 \pm 0.0007 $ \\
 660.1  -  669.5   & $ 0.0003\pm  0.0007  $ & $ -0.0003 \pm 0.0006  $ & $ 0.0001 \pm 0.0004 $ \\
 670.2  -  679.5   & $ 0.0006\pm  0.0008  $ & $ -0.0006 \pm 0.0010  $ & $ 0.0002 \pm 0.0005 $ \\
 679.9  -  689.9   & $ 0.0000\pm  0.0006  $ & $ 0.0007 \pm 0.0005  $ & $ 0.0002 \pm 0.0004 $ \\
 690.3  -  700.3   & $ 0.0012\pm  0.0005  $ & $ 0.0003 \pm 0.0012  $ & $ 0.0008 \pm 0.0006 $ \\
 \hline
\end{tabular}
\caption{{Best Fitting Transit Radius Ratios for R150 Spectral Lightcurves (Individual Observations)}}
\label{table_r150spec}
\end{table*}

\end{document}